\documentclass[12pt]{article}
\usepackage{tikz}
\usetikzlibrary{matrix,arrows,decorations.pathmorphing,decorations.pathreplacing}
\usepackage{jheppub}
\usepackage{amsmath,amssymb,euscript,array,mathrsfs,appendix,ctable,marvosym}
\usepackage{arydshln}
\usepackage{todonotes}
\usepackage{graphicx}
\usepackage[normalem]{ulem}

\def\mf{\mathfrak f} 
 
\def\mh{\mathfrak h}
\def\muu{\mathfrak u}

\def\msu{\mathfrak{su}}

\newcommand{\Tr}{\operatorname{Tr}}

\def\Ba{{\boldsymbol a}}
\def\Bb{{\boldsymbol b}}

\def\Bh{{\boldsymbol h}}

\def\Btheta{{\boldsymbol \theta}}

\def\Blambda{{\boldsymbol\lambda}}

\def\Bv{{\boldsymbol v}}
\def\Bomega{{\boldsymbol \omega}}

\def\B0{{\boldsymbol 0}}

\def\BOmega{{\boldsymbol\Omega}}
\def\Bvarpi{{\boldsymbol\varpi}}

\def\CP{{\mathbb C}P}

\def\SU{\text{SU}}
\def\U{\text{U}}

\def\Z{{\bf Z}}

\def\Z{\boldsymbol{Z}}

\newcommand{\Be}{\boldsymbol{e}}
\newcommand{\Balpha}{{\boldsymbol{\alpha}}}

\def\Dbarslash{\,\,{\raise.15ex\hbox{/}\mkern-12mu {\bar D}}}
\def\Dslash{\,\,{\raise.15ex\hbox{/}\mkern-12mu D}}
\def\delslash{\,\,{\raise.15ex\hbox{/}\mkern-9mu \partial}}
\def\delbarslash{\,\,{\raise.15ex\hbox{/}\mkern-9mu {\bar\partial}}}

\def\II{\mathscr{I}}
 
\newcommand{\Sp}{{\color{white}()}}
\def\SU{\text{SU}}
\def\SO{\text{SO}}
\def\U{\text{U}}
\def\Sp{\text{Sp}}

\newcommand{\EQ}[1]{\begin{equation}\begin{split} #1
\end{split}\end{equation}}
\newcommand{\AL}[1]{\begin{subequations}\begin{align} #1
\end{align}\end{subequations}}
\newcommand{\SP}[1]{\begin{equation}\begin{split} #1
\end{split}\end{equation}}

\title{The Structure of Non-Abelian Kinks
}
\author[a]{Timothy J. Hollowood,}
\author[b]{J. Luis Miramontes}
\author[c]{and David M. Schmidtt}
\affiliation[a]{Department of Physics, Swansea University, Swansea, SA2 8PP, U.K.}
\affiliation[b]{Departamento de F\'\i sica de Part\'\i culas and IGFAE,
Universidad
de Santiago de Compostela, 15782 Santiago de Compostela, Spain}
\affiliation[c]{Instituto de F\'\i sica Te\'orica IFT/UNESP, Rua Dr. Bento Teobaldo Ferraz 271, Bloco II, CEP 01140-070, S\~ao Paulo-SP, Brasil}

\emailAdd{t.hollowood@swansea.ac.uk}
\emailAdd{jluis.miramontes@usc.es}
\emailAdd{david.schmidtt@gmail.com}

\abstract{We consider a class of integrable quantum field theories in $1+1$ dimensions whose classical equations 
have kink solutions with internal collective coordinates that transform under a non-abelian symmetry group. 
These generalised sine-Gordon theories have been shown to be related to the world-sheet theory of the string in the AdS/CFT correspondence. We provide a careful analysis of the boundary conditions at spatial infinity complicated by the fact  that they are defined by actions with a WZ term. We go on to describe the local and non-local charges carried by the kinks
and end by showing that their structure is perfectly consistent with the exact factorizable S-matrices that have been proposed to describe these theories.
}

\notoc
\begin{document}

\maketitle

\newpage

\section{Introduction}
\label{intro}

The Symmetric Space Sine-Gordon (SSSG) theories are relativistic integrable
field theories in 1+1 dimensions whose equations of motion arise as the result of performing the Pohlmeyer reduction of a sigma model
with a symmetric space F/G as the target space~\cite{Pohlmeyer:1975nb} (for a review, see~\cite{Miramontes:2008wt} and references
therein). Their Lagrangian formulation was originally proposed in~\cite{Bakas:1995bm} in terms of the gauged Wess-Zumino-Witten (gWZW) action for a coset G/H deformed by a specific potential term. They have received recent attention because they are classically equivalent to the world-sheet theories of strings on spacetimes built in terms of symmetric spaces which are relevant in the context of the \text{AdS}/CFT correspondence~\cite{Tseytlin:2003ii,Mikhailov:2005qv,Mikhailov:2005sy,Hoare:2012nx}. 
Moreover, Pohlmeyer reduction has been generalized to the case where the symmetric space is replaced by a semi-symmetric space ${\cal F}/G$, which is the quotient of a supergroup with an ordinary group. This gives rise to another class of integrable models which include fermionic degrees of freedom and can be formulated as the (bosonic) gWZW action for a coset $G/H$ with a potential term coupled also to a set of two-dimensional fermion fields~\cite{Grigoriev:2007bu,Mikhailov:2007xr}. These theories based on a semi-symmetric space have been called the Semi-Symmetric Space Sine-Gordon (SSSSG) theories. We will refer to the general class of theories, SSSG or SSSSG, as ``generalized SG" theories.

The equivalence between world-sheet and generalized SG
theories is classical, and it does not seem possible that it can
be maintained at the quantum level in general, since the two descriptions have a different Poisson
structure~\cite{Mikhailov:2005qv,Mikhailov:2005sy,Mikhailov:2006uc,Schmidtt:2011nr}. In fact, Pohlmeyer reduction involves a specific modification of the Poisson structure of the sigma model aimed to alleviating its non-ultralocality~\cite{Delduc:2012qb,Delduc:2012vq}.
However, in the context of $\text{AdS}_5/\text{CFT}_4$, it was argued that quantum equivalence may hold~\cite{Grigoriev:2007bu,Mikhailov:2007xr} (see also~\cite{Grigoriev:2008jq,Roiban:2009vh,Hoare:2009rq,Hoare:2009fs,Iwashita:2010tg,Hoare:2010fb,Iwashita:2011ha}) so that the Lagrangian formulation of
the SSSG theory would provide the starting point to find a novel and manifestly two-dimensional
Lorentz invariant formulation of the full $\text{AdS}_5 \times S^5$ superstring theory. 

Understanding the fate of the equivalence at the quantum level requires the knowledge of the quantum solution of the generalized SG theories which, until very recently, was extremely limited. 
Actually, it included only
the (bosonic) generalized SG theories related to the symmetric spaces $S^2=\SO(3)/\SO(2)$ and $S^3=\SO(4)/\SO(3)$, which are
the well-known sine-Gordon~\cite{Zamolodchikov:1978xm} and complex sine-Gordon~\cite{Dorey:1994mg}
theories, respectively, and the homogeneous sine-Gordon theories~\cite{FernandezPousa:1996hi,FernandezPousa:1997zb,Miramontes:1999hx}, which are Pohlmeyer reductions of the principal chiral models. In all these cases the gauge symmetry group $H$ is either trivial, for the $S^2$ case, or abelian.
More recently, and motivated by their applications in string theory, the exact S-matrices for the generalized SG theories corresponding to $\CP^{n+1}=\SU(n+2)/\U(n+1)$ and to the semi-symmetric space ${\cal F}/G=\frac{P\SU(2,2|4)}{\Sp(2,2)\times \Sp(4)}$ have been explicitly constructed in~\cite{Hollowood:2010rv} and~\cite{Hoare:2011nd,Hoare:2013ysa}. In these two cases $H$ is non-abelian; namely, H= $\U(n)$ and $\SU(2)^{\times 4}$, respectively.

In order to have the full picture one needs to understand how to quantize the theories with a non-abelian symmetry group $H$ in which the solitons, or kinks, carry non-abelian internal degrees-of-freedom. For $\CP^{n+1}$, for which the quantum picture is available, the conjecture for the S-matrix is based on the semiclassical quantization of the spectrum of solitons which has been worked out for all the symmetric spaces of type~I in~\cite{Hollowood:2010dt}. However, there is an apparent mismatch between the classical soliton solutions and the set of quantum states described by the S-matrix in~\cite{Hollowood:2010rv}. The latter are kinks that interpolate between a discrete set of vacuum states identified with the irreducible representations of $\SU(n)$ of level $\leq k$, where $k$ is the level of the WZW action in the Lagrangian of the theory. They are labelled by two dominant weights $\Ba,\Bb$ of level $\leq k$ such that the topological charges $\Bb-\Ba$ are weights of the symmetric representations of the quantum group $U_q(\SU(n))$, with $q=e^{i\pi/(k+n)}$. In contrast, the semi-classical solitons are labelled just by their Noether charge and transform in symmetric representations of $H=\SU(n)$.
Similarly, for ${\cal F}/G=\frac{P\SU(2,2|4)}{\Sp(2,2)\times \Sp(4)}$ the physical (unitary) S-matrix describes the scattering between kinks that interpolate between a discrete set of vacuum states~\cite{Hoare:2013ysa}, while the corresponding solitons are labelled only by a Noether charge~\cite{Hollowood:2011fq}.

In fact this mismatch can already be appreciated at the level of the S-matrix description. The physical unitary S-matrix is obtained by performing a vertex to IRF (Interaction-Round-a-Face) transformation that is familiar from integrable models in statistical mechanics. The so-called vertex form describes states that transform in representations of $U_q(H)$ but the corresponding S-matrix is non-unitary. The transformation to the IRF form involves a change of basis along with a kind of gauge fixing and leads to a new unitary S-matrix describing a set of kinks as described above.
The aim of this paper is to show that the kink picture used in the construction of the IRF S-matrix arises in a natural way from the Lagrangian formulation of the generalized SG theories.
It follows from a careful definition of the Lagrangian action by taking into account the need of describing field configurations with non-trivial boundary conditions. Taking proper account of the boundary conditions and their implication for solitons leads to what amounts to a semi-classical realization of the vertex to IRF transformation, a point-of-view that will be amplified in the follow up paper \cite{Nextpap}.

The generalized SG theories are examples of field theories defined in terms of an action that includes a Wess-Zumino topological term whose consistency imposes two types of quantization conditions. The first one is the well known quantization of the coupling constant, whose role is taken by the level of the WZW action~\cite{Witten:1983ar}.
The second has not been considered so far in this context and is required to define the WZ term on a world-sheet with boundary. An important feature of the generalized SG theories is that they admit soliton solutions with non-trivial boundary conditions at $x\to \pm\infty$~\cite{Hollowood:2010dt,Hollowood:2011fq}. This prevents from considering the 1+1 dimensional space to be compact and forces us to define the WZ topological term on a world-sheet with boundary. 
In fact, it can be properly defined only for specific types of boundary conditions. Moreover, its consistency imposes quantization conditions on the boundary conditions themselves which arised originally in the study of D-branes in group manifolds~\cite{Alekseev:1998mc,Gawedzki:1999bq,Gawedzki:2001ye,Elitzur:2001qd,Kubota:2001ai}. 
In this paper we will consider the boundary conditions corresponding to on-shell configurations of minimal energy, which are the ones satisfied by the soliton solutions of~\cite{Hollowood:2010dt}. Then, the consistency of the WZ term imposes quantization conditions on them, and
the resulting picture is that the solitons are kinks (or open strings) that interpolate between 
a discrete set of vacuum states represented by conjugacy classes (or D-branes) of the gauge group $H$ labelled by dominant weights of $H$ of level $\leq k$. They are the semi-classical realization of the IRF excitations that appear in the unitary S-matrix. 
In addition, we will propose the improved action~\eqref{ala2} that includes two boundary terms. One of them is needed to ensure gauge invariance, while the other is required to make the action stationary for solitons solutions. Remarkably, this action coincides with the one considered in the perturbative calculations of~\cite{Hoare:2009fs} and the semiclassical quantization performed in~\cite{Hollowood:2010dt,Hollowood:2011fq,Hollowood:2011fm} only up to higher order terms in both perturbation theory and the semiclassical expansion.

The plan of the paper is as follows. In section~\ref{SSSG} we will summarize the basic features of the generalized SG theories and work out the form of the configurations of minimal energy. In particular, we will show that for the field $\gamma$ they are of, so-called, fully symmetric type. In section~\ref{cc} we review the construction of the infinite tower of conserved charges implied by integrability originally performed in~\cite{Hollowood:2010dt}. Compared to that reference, we will keep the gauge fixing prescription free which makes explicit the non-local character of some of the resulting charges and clarifies their behaviour under gauge transformations. Section~\ref{lagrangian}, which is the main part of the article, is devoted to the detailed construction of the Lagrangian action whose final form is given by~\eqref{ala2}. 
Then, in section~\ref{noether} we will identify the underlying physical symmetry group of the theory, which consists on global (vector) gauge transformations acting on a particular gauge slice, and construct the corresponding Noether charges. 
We will also show that the action is invariant under a discrete group of abelian (axial) transformations.
In Section~\ref{solitons} we apply the quantization conditions imposed by the consistency of the Lagrangian action
to the soliton solutions constructed in~\cite{Hollowood:2010dt}. We will discuss in detail the theories corresponding to $\CP^{n+1}$ and show 
that those quantization conditions agree with the semiclassical quantization performed in~\cite{Hollowood:2010dt}. In addition, we will construct a more general class of soliton solutions which are naturally described as kinks and whose structure fits nicely the kink picture used in~\cite{Hollowood:2010rv} to construct the S-matrix. 
Finally, in section~\ref{end} we draw some conclusions.

\section{Soliton Boundary Conditions}
\label{SSSG}

In this section we review the main features of the generalized SG theories.
The discussion will be focused on the case of theories constructed in terms of symmetric spaces, which means either the SSSG theories or  the SSSSG theories with the fermion fields set to zero, since they play no role in the discussion.
More details can be found in~\cite{Miramontes:2008wt,Grigoriev:2007bu} and the references therein. 
In particular, we work out the form of the configurations of minimal energy which provide the boundary conditions satisfied by the soliton solutions constructed in~\cite{Hollowood:2010dt}.

The starting point is a symmetric space realized as a
quotient of two Lie groups $F/G$.\footnote{For semi-symmetric spaces ${\cal F}/G$, $F$ is the bosonic subgroup of ${\cal F}$.} The group in the numerator $F$ admits 
an involution $\sigma_-$ whose stabilizer is the
subgroup $G$. Acting on the Lie algebra of $F$, the involution gives rise to the canonical 
decomposition
\EQ{
{\mathfrak f} = {\mathfrak g} \oplus {\mathfrak p}
\quad \text{with} \quad 
[{\mathfrak g},{\mathfrak g}]\subset {\mathfrak g}\>, 
\quad [{\mathfrak g},{\mathfrak p}]\subset 
{\mathfrak p}\>, 
\quad [{\mathfrak p},{\mathfrak p}]\subset {\mathfrak g}\>,
\label{CanonicalDec}
}
where ${\mathfrak g}$ and ${\mathfrak p}$ are the $+1$ and $-1$ eigenspaces of $
\sigma_-$, respectively.
In this paper we will consider compact symmetric spaces of type I, which are those for which $F$ is a compact simple Lie group. 
Moreover, just for simplicity, we will restrict ourselves to symmetric spaces of rank one, which means that the maximal abelian subspaces of ${\mathfrak p}$ are of dimension one. Particular examples are the spheres $S^n=\SO(n+1)/\SO(n)$ and the complex projective spaces $\CP^{n+1}=\SU(n+2)/\U(n+1)$.

The equations of motion are formulated at the level of the Lie algebra $\mathfrak f$
and involve two fields $\gamma(t,x)\in G$ and $A_\mu(t,x)$. They can be written as a zero-curvature condition for a connection that depends on an auxiliary spectral parameter $z$
\EQ{
{\cal L}_\mu = \partial_\mu +{\cal A}_\mu\,,\qquad 
[{\cal L}_\mu,{\cal L}_\nu]=0\,,
\label{zc}
}
where\footnote{In our notation, $x^\pm  = t\pm x$ are light-cone coordinates and for a general 2-vector we use $a_\pm=\frac12(a_0\pm a_1)$. Our choice of metric is $\eta=\text{diag}(1,-1)$ and we normalize the anti-symmetric symbol with $\epsilon_{01}=1$.
Then, $\eta^{\mu\nu}a_\mu b_\nu= 2(a_+b_-+a_-b_+)$ and $\epsilon^{\mu\nu}a_\mu b_\nu= 2(a_+b_--a_-b_+)$.
}
\EQ{
&
{\cal L}_+=\partial_++\gamma^{-1}\partial_+\gamma+\gamma^{-1}A_+\gamma-z\mu \Lambda\,,\\[5pt]
&
{\cal L}_-=\partial_-+A_- -z^{-1}\mu\gamma^{-1}\Lambda\gamma\,.
\label{zc2}
}
Here, $\Lambda$ is a constant element of a maximal abelian subspace of  ${\mathfrak p}$, $\mu$ is a mass scale, and it is straightforward to check that the zero-curvature condition $[{\cal L}_\mu,{\cal L}_\nu]=0$ is independent of the value of $z$. The adjoint action of 
 $\Lambda$ gives rise to the orthogonal decomposition
 \EQ{
{\mathfrak f} ={\mathfrak f}^\perp\oplus{\mathfrak f}^\parallel\ ,\qquad
{\mathfrak f}^\perp=
 \mathop{\rm Ker} {\mathop{\rm Ad}}_{\Lambda}\ , \qquad{\mathfrak f}^\parallel = \mathop{\rm Im} {\mathop{\rm Ad}}_{\Lambda}
\label{Orthogonal}
}
that, schematically, satisfies
\SP{
[{\mathfrak f}^\perp,{\mathfrak f}^\perp]\subset {\mathfrak f}^\perp\ ,\qquad
[{\mathfrak f}^\perp,{\mathfrak f}^\parallel]\subset{\mathfrak f}^\parallel\,.
\label{Orthogonal2}
}
A central role is played by the subgroup $H\subset G$ that keeps $\Lambda$ fixed under adjoint action. Namely, 
$H=\{h\in G \mid h\Lambda h^{-1} =\Lambda\}$ so that the Lie algebra of $H$, denoted by ${\mathfrak h}$, consists of the elements of ${\mathfrak g}$ that commute with $\Lambda$. Then, the fields $A_\pm\in{\mathfrak h}$ are the light-cone components of a gauge field
associated to the gauge symmetry transformations
\EQ{
\gamma\to h\gamma h^{-1}\ ,\qquad 
A_\mu\to h\big(A_\mu+\partial_\mu\big)h^{-1}\ ,\qquad
h\in H\ .
\label{gauge}
}
The field $A_\mu$ satisfies the constraints 
\SP{
\Big(\gamma^{\mp1}\partial_\pm\gamma^{\pm1}
+\gamma^{\mp1}A_\pm\gamma^{\pm1}\Big)^\perp=A_\pm
\label{gco2}
}
which, introducing the covariant derivative $D_\mu\gamma=[\partial_\mu+A_\mu,\gamma]$, can be written in the more compact form
\EQ{
\left(\gamma^{-1}D_+\gamma\right)^\perp=\left(D_-\gamma\gamma^{-1}\right)^\perp=0\,.
\label{gco3}
}

An important feature of the generalized SG theories is that they admit soliton solutions with non-trivial boundary conditions at spatial infinity, $x=\pm\infty$. 
Classically, the vacuum is degenerate and there is a space of on-shell vacuum configurations of minimal energy given by covariantly constant group elements $\gamma\in H$. This can be shown by considering the  density of energy, which can be written as~\cite{Hollowood:2010dt} (see~\eqref{enermom})
\EQ{
T_{00}=-\frac{\kappa}{4\pi}\Tr\left[\left(\gamma^{-1}D_+\gamma\right)^2 +\left(\gamma^{-1}D_-\gamma\right)^2-4\mu^2\Lambda\gamma^{-1}\Lambda\gamma\right]\,.
}
Since $G$ is compact, the configurations of minimal energy correspond to
\EQ{
D_\mu\gamma=0 \,,\qquad \gamma \in H \,.
\label{me}
}
On-shell, the gauge field is flat, $[\partial_\mu +A_\mu,\partial_\nu+A_\nu]=0$, and it can be written as \mbox{$A_\mu=-\partial_\mu U U^{-1}$} with $U\in H$. Then, the configurations of minimal energy turn out to be of the form
\EQ{
\gamma^{\text{vac}}= {U^\text{vac}}\, f \, {U^\text{vac}}^{-1}\,,\qquad A_\mu^{\text{vac}}=-\partial_\mu {U^\text{vac}}\, {U^\text{vac}}^{-1}
\label{vac1}
}
with $f\in H$ constant.
Notice that ${U^\text{vac}}\in H$ is a Wilson line
\EQ{
{U^\text{vac}}=\text{P}\,\exp\Big[-\int_{x_0}^{x} dx^\mu A_\mu^\text{vac}\Big]\equiv {U^\text{vac}}(x;x_0)
\label{Wilson1}
}
that depends on an arbitrary reference point $x_0$ so that $f=\gamma^\text{vac}(x_0)$. This exhibits that $f$ is constant but not gauge invariant. Instead, under the gauge transformations~\eqref{gauge},
\EQ{
{U^\text{vac}}(x;x_0) \to h(x) {U^\text{vac}}(x;x_0) h^{-1}(x_0)\,,\qquad
f\to h(x_0) f h^{-1}(x_0)\,.
\label{constant}
}
The vacuum configurations~\eqref{vac1} are gauge equivalent to the configurations where $\gamma=f$ is constant and $A_\mu=0$, which are the vacuum configurations considered in~\cite{Hollowood:2010dt}. They provide the boundary conditions of the soliton solutions constructed in that article where
the gauge was fixed by imposing $A_\mu=0$ at the level of the equations of motion. 
Off-shell, we shall consider boundary conditions corresponding to the configurations of minimal energy~\eqref{vac1}, but leave the gauge fixing prescription free.
Namely, 
\EQ{
\gamma\big|_B = UfU^{-1}\big|_B\,,\qquad 
A_\mu\big|_B=-\partial_\mu UU^{-1}\big|_B
\,.
\label{bc1}
}
Taking into account~\eqref{Wilson1}, the explicit  form of the field $\gamma$ on the boundary is
\EQ{
&\gamma(t,\pm\infty)=U_{t_0}(t,\pm\infty)\, f_{\pm}\, U_{t_0}^{-1}(t,\pm\infty)\,,\\[5pt]
&
U_{t_0}(t,\pm\infty)= \text{P}\,\exp\Big[-\int_{t_0}^{t} d\tau\, A_0(\tau,\pm\infty)\Big]\,,\qquad f_\pm = \gamma(t_0,\pm\infty)\,,
\label{bcgood}
}
where $t_0$ is arbitrary.
These boundary conditions respect the gauge symmetry~\eqref{gauge}, so that if $\gamma$ is allowed on the boundary then the symmetry implies
that $h\gamma h^{-1}$ should also be allowed for any $h\in H$. Therefore, on the boundary, the field $\gamma$ will be allowed to take values in the whole conjugacy class, or co-adjoint orbit,\footnote{For compact semi-simple Lie groups the adjoint orbits are the same as the co-adjoint orbits}
$C_f(H)=\bigl\{ U f U^{-1}\mid U\in H\bigr\}$ labelled by the constant element $f\in H$. This will be one of the key ingredients to construct the Lagrangian action of the generalized SG theories in section~\ref{lagrangian}.

The conjugacy classes specified by two conjugated constant elements $f$ and $hfh^{-1}$ of $H$ are identical, and it is useful to recall that any element of a compact Lie group $H$ can be conjugated to a given maximal torus $T\subset H$. The Lie algebra of $H$ is the direct sum of a semi-simple Lie algebra and 
and abelian Lie algebra ($\mh$ is reductive); namely, $\mh=\mh^\zeta \oplus \mh_\text{ss}$ with $\mh^\zeta =\muu(1)^{\oplus p}$.  Correspondingly, the elements of $H$ can be written as a product of a component in $H^\zeta$ and a component in $H_\text{ss}$, although the decomposition is unique only up to multiplication by the elements of $H_\text{ss}\cap H^\zeta$ which is the (discrete) centre of $H_\text{ss}$.
For example, for \mbox{$\CP^{n+1}=\SU(n+2)/\U(n+1)$}, the symmetry group is $H=\U(n)=(\U(1)\times \SU(n))/ {\mathbb Z}_{n}$. Similarly, for $S^n=\SO(n+1)/\SO(n)$ it is $H=\SO(n-1)$.
Then, it is convenient to write the elements of $T\subset H$ as $\exp(2\pi i \lambda/\kappa)$, where $\lambda = \Blambda\cdot\Bh + \lambda^\zeta$ is the sum of a Cartan element $\Blambda\cdot\Bh$ of $\mh_\text{ss}$, written in terms of a basis for the Cartan subalgebra, and $\lambda^\zeta\in\mh^\zeta$. 
Furthermore, since Weyl transformations act on the Cartan subalgebra of $\mh_\text{ss}$ as $\Blambda \to h\Blambda h^{-1}$, we can restrict the inequivalent choices of $\Blambda$ to those in the fundamental Weyl chamber, which satisfy $\Blambda\cdot\Balpha\geq0$ for any positive root $\Balpha \in \Phi^+$ of $\mh_\text{ss}$. 
Then, a more explicit form of the conjugacy class generated by $f=\exp(2\pi i \lambda/\kappa)$ with $\lambda = \Blambda\cdot\Bh + \lambda^\zeta$ is
\EQ{
C_f(H)= 
\{e^{2\pi i\lambda^\zeta/\kappa}\,\, g\,  e^{2\pi i\Blambda\cdot\Bh/\kappa}\, g^{-1}\mid g\in H_\text{ss}\}\equiv \{e^{2\pi i\lambda^\zeta/\kappa}\}\times
{\cal C}_{\Blambda}(H_\text{ss})\,.
\label{coexp}
}
Namely, a point in $H^\zeta$  labelled by $\lambda^\zeta$ times a conjugacy class of $H_\text{ss}$ labelled by $\Blambda$. Taking also into account that $H_\text{ss}$ is compact, there is a different conjugacy class ${\cal C}_{\Blambda} (H_\text{ss})$ for each $\Blambda$ in the {\em classical moduli space}
\EQ{
{\cal M}_{\text{cl}}= \{\Blambda\mid 0\leq \Blambda\cdot\Balpha \leq \kappa\,,\; \forall\,\Balpha\in \Phi^+\}\,.
\label{clms}
}
Notice that the dimension of ${\cal C}_{\Blambda} (H_\text{ss})$ depends on $\Blambda$. For instance, for $H=\SU(2)$, which is isomorphic to the 3-sphere $S^3$, the coadjoint orbits are 2-spheres for $0< \Blambda\cdot\Balpha < \kappa$ and points for $\Blambda\cdot\Balpha =0,\kappa$, the latter corresponding to the two elements in the (discrete) centre of $\SU(2)$.

\section{Integrability and Conserved Charges}
\label{cc}

In~\cite{Hollowood:2010dt}, the infinite tower of conserved charges implied by integrability was written in terms of a subtracted monodromy that was constructed using the on-shell gauge fixing condition $A_\mu=0$. For our purposes, it will be convenient to write them leaving the gauge fixing prescription free and taking into account that vacuum states are of the form~\eqref{vac1}. 
However, the subtracted monodromy will still be constructed on-shell, which in particular means that $A_\mu$ is flat and can be written as
\EQ{
A_\mu= -\partial_\mu UU^{-1}\,.
\label{flat}
}
Here, $U\in H$ is a Wilson line
\EQ{
U=\text{P}\,\exp\Big[-\int_{x_0}^{x} dx^\mu A_\mu\Big]\equiv U(x;x_0)
\label{Udef}
}
that depends non-locally on $A_\mu$ and on an arbitrary reference point $x_0$. Its precise definition requires a one-dimensional curve going from $x_0$ to $x\equiv(t,x)$ whose choice is irrelevant because $A_\mu$ is flat. Under the gauge symmetry~\eqref{gauge}, the transformation of $U$ is
\EQ{
U(x,x_0) \to h(x) U(x,x_0) h^{-1}(x_0)\,.
\label{gaugeUa}
}

We start with the solution to the associated linear problem
\EQ{
{\cal L}_\mu(z) \Upsilon(t,x;z)=0\,,
\label{alp}
}
whose integrability conditions are the equations of motion~\eqref{zc}.
Then, we define the subtracted monodromy
\EQ{
{\cal M}(z)= \lim_{x\to\infty} U^{-1}(x) \Upsilon_0^{-1}(x;z) \Upsilon(x;z)\Upsilon^{-1}(-x;z)\Upsilon_0(-x;z)U(-x)\,,
\label{mono}
}
where
\EQ{
\Upsilon_0(x;z)= \exp\big[(z x^+ + z^{-1} x^-)\mu\Lambda\big]
\label{gamma0}
}
and we have omitted the dependence on $t$ and on $x_0$ to make the notation lighter.
Using~\eqref{flat} and~\eqref{alp}, it can be easily shown that
\EQ{
\partial_0\big(U^{-1}\Upsilon_0^{-1}\Upsilon\big)= -U^{-1}\big(\gamma^{-1}D_+\gamma + z^{-1}\mu\big(\Lambda-\gamma^{-1}\Lambda\gamma)\big)\Upsilon_0^{-1}\Upsilon\,.
}
Therefore, with the boundary conditions~\eqref{bc1}, it follows that ${\cal M}(z)$ is conserved: \mbox{$\partial_0{\cal M}(z)=0$}.
In addition, since the solution to~\eqref{alp} for the vacuum configurations~\eqref{vac1} is
\EQ{
\Upsilon^{\text{vac}}(x;z)= U(x)\Upsilon_0(x;z) \alpha(z)
}
with $\alpha(z)\in H$ independent of~$t$ and~$x$, the corresponding value of the subtracted monodromy is ${\cal M}^{\text{vac}}(z)=1$. 
Finally, it is important to notice that the subtracted monodromy is not fully gauge invariant as a consequence of its implicit dependence on the reference point $x_0$ introduced in~\eqref{Udef}.
This can be easily checked by noticing that, under~\eqref{gauge},
\EQ{
\Upsilon(x;z) \Upsilon^{-1}(y;z)\to h(x)\, \Upsilon(x;z) \Upsilon^{-1}(y;z) \, h^{-1}(y)
}
which, taking~\eqref{gaugeUa} into account, leads to
\EQ{
{\cal M}(z) \to h(x_0) {\cal M}(z) h^{-1}(x_0)\,.
\label{nonlocgauge}
}
Therefore, the subtracted monodromy is invariant under 
the gauge transformations that satisfy $h(x_0)=1$ which exhibits that it is a non-local object that depends implicitly on the reference point $x_0$. Notice that global gauge transformations satisfy $h(x_0)\not=1$, but the converse is not true; namely, $h(x_0)\not=1$ is not enough to ensure that the transformation is global.
The non-local nature of the monodromy will imply later that the soliton/kinks of the theory carry non-local charges under the non-abelian part of the symmetry group in a way that will be made precise. This is an important feature because it is well know (for example in the work of Bernard and LeClair~\cite{Bernard:1990ys}) that the associated non-local conserved currents in the quantum theory have non-trivial monodromies which lead to the quantum charges satisfying a quantum group deformation of the Lie algebra of the symmetry group rather than the conventional Lie algebra.

The
expansion of the subtracted monodromy around $z=0$ and~$\infty$
\EQ{
{\cal M}(z)=\exp\big[
q_0+q_{1}z+q_{2}z^2+\cdots\big]=\exp\big[q_{-1}/z+q_{-2}/z^2+\cdots\big]
\label{mex}
}
provides a set of  conserved charges $q_s$ of Lorentz spin $s$, and we will show that $q_s\in\mf^\perp$.
Taking~\eqref{nonlocgauge} into account, their change under gauge transformations is
\EQ{
q_s\to h(x_0)\, q_s\, h^{-1}(x_0)\,.
\label{qtransf}
}
Therefore, the projection of $q_s$ onto the centre of ${\mathfrak f}^\perp$ is gauge invariant. Moreover, it can be shown that these gauge invariant conserved charges provide an infinite number of local conserved charges.

The explicit form of the  conserved charges can be deduced using the Drinfeld-Sokolov procedure~\cite{DeGroot:1991ca} following the approach of~\cite{Hollowood:2010dt}, which we briefly summarize in the following in order to specialize it to our case. 
First of all, recall that the proper algebraic setting for the Lax connection~\eqref{zc2} is the affine (loop) Lie algebra with a gradation that is fixed by the decomposition~\eqref{CanonicalDec}:
\EQ{
\hat{\mathfrak  f}= \bigoplus_{n\in\Z} \left(z^{2n} \otimes {\mathfrak g}
  +z^{2n+1} \otimes {\mathfrak p} \right)\equiv\bigoplus_{k\in
  \boldsymbol{Z}}\>\hat{\mathfrak f}_k\ ,
}
where we have defined
\SP{
\hat{\mathfrak f}_k = \begin{cases}
z^{k}\otimes \mathfrak{g}\,,      & \text{if}\;\; k=2n\,, \\
z^{k}\otimes \mathfrak{p}\,,      & \text{if}\;\; k=2n+1\,,
\end{cases}
}
and $[\hat{\mathfrak f}_k,\hat{\mathfrak f}_l]\subset \hat{\mathfrak f}_{k+l}$.
In the following we will often use the notation
$
\hat{\mathfrak f}_{<0}=\bigoplus_{k<0}\hat{\mathfrak f}_k$, $\hat{\mathfrak f}_{\geq0}=\bigoplus_{k\geq0}\hat{\mathfrak f}_k$,
etc.

We start by considering the charges of positive spin and introduce\EQ{
\Phi=\exp\,y(z)\>,\qquad y(z)=\sum_{s\geq1}z^{-s}y_{-s} \in \hat{\mathfrak f}_{<0}\,,
}
and solve
\EQ{
\Phi^{-1} {\cal L}_+(z) \Phi=  \partial_+  - z\mu\Lambda+ h_+(z)\ ,\qquad
h_+(z)=\sum_{s\geq0} h_{-s,+} z^{-s}\in\hat{\mathfrak f}^\perp_{\leq0} \ .
\label{DS}}
Correspondingly, using the zero curvature condition~\eqref{zc},
\EQ{
\Phi^{-1} {\cal L}_-(z)\Phi = \partial_-  +h_-(z)\ ,
\qquad
h_-(z)=\sum_{s\geq0}h_{-s,-}z^{-s}\in \hat{\mathfrak f}^\perp_{\leq0}\ .
\label{DS2}
}
In these equations $\Phi$ can always be choosen such that $\Phi$ and $h_\pm(z)$ are local functions of the component fields by simply enforcing the condition
\EQ{
y(z)\in  \hat{\mathfrak f}^\parallel_{<0}\,.
\label{LocalCond}
}
More precisely, $\Phi$ and $h_+(z)$ are local functions of the combination of fields 
\EQ{
{\cal L}_+(z)-\partial_++ z\mu\Lambda=\gamma^{-1}\partial_+\gamma +\gamma^{-1}A_+\gamma= \gamma^{-1} D_+\gamma + A_+
}
and, since for vacuum configurations $\gamma\in H$ and $D_\pm\gamma=0$,
\EQ{
\Phi^{\text{vac}}=1\,,\qquad h_+^{\text{vac}}= A_+^{\text{vac}}\,,\qquad h_-^{\text{vac}}= A_-^{\text{vac}} -z^{-1}\mu\Lambda\,.
}
The explicit expression of the densities of spin~1 and~2 can be found in~\cite{Hollowood:2010dt}. Namely, for spin~1
\EQ{
h_{0,\pm}=A_\pm\,,
\label{spin1}
}
while the densities of spin~2 provide, in particular, the components of the stress-energy tensor
\EQ{
&\Tr\big(\Lambda h_{-1,+}\big)\sim T_{++}=-\frac{\kappa}{4\pi}\Tr\Big[\bigl(\gamma^{-1}D_+\gamma\bigr)^2\Bigr]
\\[5pt]
&\Tr\big(\Lambda h_{-1,-}\big)\sim -T_{-+}=\frac\kappa{2\pi}\,\mu^2\,{\rm Tr}\Big[\Lambda\gamma^{-1}\Lambda\gamma
\Big]
\label{enermom}
}
which are gauge invariant. In terms of $h_+$ and $h_-$, the zero curvature condition reads
\EQ{
\bigl[\partial_+ + h_+(z),  \partial_-  + h_-(z)\bigr]=0
\label{Zero2}
}
which proves directly that the projection of $h_\pm(z)$ onto ${\mathfrak z}(\Lambda)$, the centre of $\hat{\mathfrak f}^\perp$, leads to {\em local\/} conserved currents. Since ${\mathfrak z}(\Lambda)$ always contains
the infinite set of elements $z^{2n+1}\Lambda$,
as well as the abelian factor $\mh^\zeta=\muu(1)^{\oplus p}$ of ${\mathfrak h}$ times $z^{2n}$, 
there are infinite local conserved charges of positive spin.
In a similar way, the set of conserved quantities with negative spin can be constructed starting from
\EQ{
\gamma{\cal L}_-(z)\gamma^{-1}&= \partial_- -\partial_-\gamma\gamma^{-1}+ \gamma A_-\gamma^{-1} - z^{-1}\mu\Lambda\ ,\\
\gamma{\cal L}_+(z)\gamma^{-1}&=\partial_++A_+-z\mu\gamma\Lambda\gamma^{-1}
\label{DSMinus}
}
instead of ${\cal L}_\pm$, with 
\EQ{
\Phi\rightarrow \tilde\Phi\in \exp \hat{\mathfrak f}^\parallel_{>0}\ ,\qquad
h_\mu(z)\rightarrow \tilde h_\mu(z)\in \hat{\mathfrak f}^\perp_{\geq0}\ .
}
and
\EQ{
\tilde h_\mu(z)=\sum_{s\geq 0} h_{s,\mu}z^s\ .
}
Both constructions, and in particular the quantities $h_\mu(z)$ and $\tilde h_\mu(z)$, are trivially related by means of the replacements
\EQ{
z\rightarrow z^{-1},\quad
\partial_+\rightarrow \partial_-,\quad
\gamma\rightarrow \gamma^{-1},\quad
A_\pm\rightarrow A_\mp\ .
\label{Parity}
}
Therefore, for vacuum configurations,
\EQ{
\tilde\Phi^{\text{vac}}=1\,,\qquad \tilde h_+^{\text{vac}}= A_+^{\text{vac}}-z\mu\Lambda \,,\qquad \tilde h_-^{\text{vac}}= A_-^{\text{vac}} \,.
}

The next step is to solve the zero curvature~\eqref{Zero2}  as follows
\EQ{
h_+(z)= \Omega\partial_+ \Omega^{-1}\>,\qquad
h_-(z)= -z^{-1}\mu\Lambda+ \Omega\partial_- \Omega^{-1}\>,
\qquad
\Omega\in \exp \hat{\mathfrak f}^\perp_{\leq0}\,,
}
which leads to
\EQ{
\chi^{-1}{\cal L}_\pm(z)\chi=\partial_\pm -z^{\pm1} \mu\Lambda\>,\qquad
\chi = \Phi\Omega \in \exp \hat{\mathfrak f}_{\leq0}\ .
\label{MiniLax1}
}
In other words, $\chi=\chi(z)$ is a formal series in $z^{-1}$ taking values in $F$. This provides the following expression for the solution to the associated linear problem~\eqref{alp}
\EQ{
\Upsilon(z)= \chi(z) \Upsilon_0(z) g_+(z),
\label{LPsol1}
}
where $\Upsilon_0(z)$ is given by \eqref{gamma0} and
$g_+(z)$ is a constant element of the loop group.
In a completely analogous fashion, starting from $\gamma{\cal L}_-(z)\gamma^{-1}$ instead of ${\cal L}_+(z)$ we get
\EQ{
\tilde\chi^{-1}\gamma{\cal L}_\pm(z)\gamma^{-1}\tilde\chi=\partial_\pm -z^{\pm1} \mu\Lambda\>,\qquad
\tilde\chi=  \tilde\Phi\tilde\Omega\in \exp \hat{\mathfrak f}_{\geq0}\>,
\label{MiniLax2}
}
where
\EQ{
\tilde h_+(z)= -z\mu\Lambda +\tilde \Omega\partial_+ \tilde\Omega^{-1}\>,\qquad
\tilde h_-(z)= \tilde\Omega\partial_-\tilde\Omega^{-1}\>.
}
In this case, $\tilde\chi=\tilde\chi(z)$ is a formal series in $z$. This provides a second expression for the solution to the associated linear problem
\EQ{
\Upsilon(z)= \gamma^{-1} \tilde\chi(z) \Upsilon_0(z) g_-(z)\ ,
\label{LPsol2}
}
where $g_-(z)$ is another constant element of the loop group. Equating~\eqref{LPsol1} and~\eqref{LPsol2} gives rise to the factorization (Riemann-Hilbert) problem
\EQ{
\Upsilon_0(z) g_-(z)g_+(z)^{-1} \Upsilon_0^{-1}(z)=\tilde\chi(z)^{-1}\gamma\chi(z)\>.
\label{RHp}
}
Since with our boundary conditions
\EQ{
\lim_{x\to\pm\infty}\Phi(x;z)=1\ ,\qquad
\lim_{x\to\pm\infty}\tilde\Phi(x;z)=1\,,
}
eqs.~\eqref{LPsol1} and~\eqref{LPsol2} provide two alternative expressions for the subtracted monodromy~\eqref{mono}
\AL{
{\cal M}(z)&=
U^{-1}(\infty)\text{P}\,\exp\,\left[-\int_{-\infty}^{+\infty}dx\, \big(h_1(z)-z^{-1}\mu\Lambda\big)\right]U(-\infty)
\label{ExpInftyB}\\[5pt]
&=U^{-1}(\infty)\gamma^{-1}(\infty)\> \text{P}\,\exp\,\left[-\int_{-\infty}^{+\infty}dx\, \big(\tilde h_1(z)+z\mu\Lambda\big)\right]\gamma(-\infty)U(-\infty)\>.\label{ExpZeroB}
}
Evaluating \eqref{ExpInftyB} at $z=\infty$ confirms the normalization of \eqref{mex}
\EQ{
{\cal M}(\infty)=U^{-1}(\infty)\> \text{P}\,\exp\,\left[-\int_{-\infty}^{+\infty}dx\, h_{0,1}\right]U(-\infty)=1\,,
}
where we have used~\eqref{spin1}
\EQ{
h_{0,\mu}=\tilde h_{0,\mu}= A_\mu=-\partial_\mu UU^{-1}\,.
}
Similarly, evaluating~\eqref{ExpZeroB} at $z=0$ gives directly the spin-zero charge of a configuration with boundary conditions~\eqref{bcgood}~\footnote{With the gauge fixing condition $A_\mu=0$, this equation simplifies to $e^{q_0}= \gamma^{-1}(\infty) \gamma(-\infty)$ which is the expression for the conserved charge $q_0$ quoted in~\cite{Hollowood:2010dt}.}
\EQ{
e^{q_0}&= {\cal M}(0)=U^{-1}(\infty)\gamma^{-1}(\infty)U(\infty)\, U^{-1}(-\infty)\gamma(-\infty)U(-\infty)\\[5pt]
&\equiv U^{-1}\big((t,\infty);x_0)\,\gamma^{-1}(t,\infty)U\big((t,\infty);(t,-\infty)\big)\gamma(t,-\infty)U\big((t,-\infty);x_0)\,,
\label{q0}
}
where in the second equation we have made explicit the dependence on $t$ and $x_0$. Its gauge transformation is provided by~\eqref{qtransf}. The Lagrangian interpretation of $q_0$ as a Noether charge will be clarified 
in section~\ref{noether} (see eq.~\eqref{NC2}).

\section{Lagrangian Formulation and Boundary Conditions}
\label{lagrangian}

The Lagrangian formulation of the SSSG theories (or the SSSSG theories with the fermion fields set to zero) was originally proposed by Bakas, Park and Shin in~\cite{Bakas:1995bm}. It is provided by the action
\SP{
S[\gamma,A_\mu]&=S_\text{gWZW}[\gamma,A_\mu]- \frac {\kappa\mu^2}{\pi} \int d^2x\,\Tr\left(\Lambda
\gamma^{-1}\Lambda\gamma\right)
\label{ala}
}
where $S_\text{gWZW}[\gamma,A_\mu]$ is the gauged WZW action for $G/H$ with coupling constant $\kappa$, so that the equations~\eqref{zc} and~\eqref{gco3} follow as the 
equations-of-motion of $S$ (see also~\cite{Miramontes:2008wt,Grigoriev:2007bu}). This action includes a Wess-Zumino topological term whose consistency at the quantum level imposes the well known quantization of the coupling constant $\kappa$. However, in order to describe field configurations with non-trivial boundary conditions, the Lagrangian action has to be formulated on a world-sheet with boundary. This requires a particular definition of  the Wess-Zumino term that takes into account the boundary conditions satisfied by the field $\gamma$ and the introduction of specific boundary terms. Before going through the details, we summarize the main features of the resulting action given by~\eqref{ala2}
\begin{itemize}
\item[i)] 
The WZ term for world-sheets with boundary, given by~\eqref{WZWb}, depends on the form of the components of $\gamma\big|_{\partial\Sigma}$ in $H_\text{ss}$, the semi-simple subgroup of $H$. Then, an important result is that its consistency at the quantum level imposes quantization conditions on the boundary conditions in addition to the well known quantization of the coupling constant. This leads to a natural description of the soliton solutions as kinks whose boundary conditions take values in topologically quantized conjugacy classes of $H_\text{ss}$.

\item[ii)]
The consistency of the action does not require the quantization of the components of $\gamma\big|_{\partial\Sigma}$ in $H^\zeta$, the abelian subgroup of $H$. However, the quantization of the components in  $H_\text{ss}$ usually implies the quantization of the components in $H^\zeta$ (see section~\ref{solitons} for an example). 

On more general grounds, the quantization of the components of $\gamma\big|_{\partial\Sigma}$ in $H^\zeta$ follows from the breakdown of the symmetry of the action under global axial $H^\zeta$ transformations which, as shown in~\eqref{anomaly}, becomes anomalous in the presence of the boundary.

\item[iii)]
In addition to the terms needed to define the Wess-Zumino term, we include two boundary terms in the action. The first one, given by~\eqref{gWZW2}, depends on the component of the gauge field $A_\mu$ in $\mh^\zeta$ and is required to ensure gauge invariance.
The second, given by~\eqref{aaaB}, amounts to a non-minimal definition of the Wess-Zumino term and is included to make the action sensitive to the boundary conditions satisfied by the gauge field $A_\mu$.

\end{itemize}

\subsection{The action on a world-sheet without boundary}

When the theory is formulated on a world-sheet $\Sigma$ without boundary, or the fields satisfy trivial boundary conditions, the gauged WZW action is
\EQ{
&
S_\text{gWZW}[\gamma,A_\mu] \\[5pt]
&
\hspace{0.5cm}
=S_\Sigma[\gamma]-\frac\kappa{\pi}\int_\Sigma d^2x\,\Tr\,\Big[
A_+\partial_-\gamma\gamma^{-1} -A_-\gamma^{-1}\partial_+\gamma-\gamma^{-1}A_+\gamma A_-+A_+A_-\Big]\,,
\label{gWZW}
}
where
\EQ{
&
S_\Sigma[\gamma]=\frac{\kappa}{8\pi} \left(\int_\Sigma d^2x\, L^\sigma(\gamma) +\frac{2}{3} \int_B d^3 x\, \omega^\text{WZ}(\tilde \gamma)\right)\,,\\[5pt]
&
L^\sigma(\gamma)=\Tr\left(\partial_\mu \gamma\, \partial^\mu \gamma^{-1}\right)\,,\qquad
\omega^\text{WZ}(\gamma)=\epsilon^{abc}\Tr \left( \gamma^{-1}\partial_a  \gamma   \gamma^{-1}\partial_b  \gamma   \gamma^{-1}\partial_c  \gamma \right)
\label{WZW}
}
is the Wess-Zumino-Witten action corresponding to the Lie group $G$.
The first term in $S_\Sigma$ is the standard sigma model action of the field $\gamma:\Sigma\to G$. The second is the Wess-Zumino topological term. It involves a three-manifold  $B$ bounded by $\Sigma$ ($\partial B=\Sigma$) and an extension $\tilde{\gamma}$ of $\gamma$ from $\Sigma$ to $B$ ($\tilde\gamma|_\Sigma=\gamma$). 
In general, the value of the Wess-Zumino term depends on $\tilde \gamma$, which makes its definition ambiguous.
At the classical level this ambiguity is not relevant because it does not affect the equations of motion. However, at the quantum level the path integral measure $\exp(iS_\text{gWZW})$ has to be independent of the choice of $\tilde \gamma$, and the WZ term has to be uniquely defined modulo $2\pi{\mathbb Z}$ for each $\gamma$. 
This imposes the well known quantization of the coupling constant $\kappa$.
Namely, if $G$ is connected and simple and the trace $\Tr$ is  normalized such that the long roots of $G$ have length squared~2, then $\kappa$ has to be an integer or half an integer~\cite{Witten:1983ar,Felder:1988sd}. In particular, 
\EQ{
\kappa=\begin{cases} k\,, & \text{for}\;\;  G=\SU(n)
\\ k/2\,, & \text{for}\;\;  G=\SO(n),\,\, \Sp(n)\ ,
\end{cases}
\label{Level}
}
where $k$ is the level of the WZW action, which is a positive integer.
The actions~\eqref{gWZW} and, hence,~\eqref{ala}  are invariant under the gauge transformations~\eqref{gauge}
\EQ{
\gamma \to h\gamma h^{-1}\,,\qquad A_\mu\to h\left(A_\mu+\partial_\mu\right)h^{-1}\,,\qquad h\in H\,.
}
Moreover, the change of the action under infinitesimal variations of the fields reads
\EQ{
&\delta S[\gamma,A_\mu]=-\frac{\kappa}{\pi}\int_\Sigma d^2x\, \Tr\Big(\delta A_+\, D_-\gamma\gamma^{-1} - \delta A_-\, \gamma^{-1}D_+\gamma\\[5pt]
&\hspace{1cm}
+\gamma^{-1}\delta\gamma\, [\partial_++\gamma^{-1}\partial_+\gamma+\gamma^{-1}A_+\gamma-z\mu \Lambda,\partial_-+A_- -z^{-1}\mu\gamma^{-1}\Lambda\gamma] \Big)
\label{gwzweom1}
}
which provides the equations of motion~\eqref{zc} and~\eqref{gco3}.

\subsection{The WZ term for soliton boundary conditions}

However, since the generalized SG theories admit soliton solutions with non-trivial boundary conditions at $x=\pm\infty$, they have to be formulated on a world-sheet with boundary. Then, there is no three-manifold $B$ such that $\Sigma=\partial B$ and the definition of the Wess-Zumino topological term 
has to be modified. In fact, it can be properly defined only for specific types of boundary conditions for the field $\gamma$ and, moreover, its consistency imposes quantization conditions on the boundary conditions themselves~\cite{Alekseev:1998mc,Gawedzki:1999bq,Gawedzki:2001ye}. For the $G/H$ gauged WZW action, the most studied class of allowed boundary conditions are the, so-called, fully symmetric ones~\cite{Elitzur:2001qd,Kubota:2001ai}
\EQ{
\gamma\big|_B = glg^{-1}\, UfU^{-1}\,,
}
where $g=g(x)\in G$, $U=U(x)\in H$, and $l\in G$ and $f\in H$ are constant.
Our boundary conditions~\eqref{bc1} are precisely of this type with $l=1$.

In the following, we will imagine the world-sheet to be a large cylinder 
\EQ{
\Sigma =S^1_{T}\times \left[ -L,+L\right]
\label{ws}
}
with Minkowskian signature. We will consider time as being periodic with period $T$ and impose non-trivial boundary conditions at $x=\pm L$. Finally, we will take the limits $T,L\rightarrow \infty $ to recover the usual 1+1 Minkowski space. Then, the boundary of $\Sigma$ consists of two timelike circles $S_\pm^1$ located at $x=\pm L$, and we can construct a two-manifold $\Sigma'$ without boundary from $\Sigma$ by gluing two disjoint disks ${\cal D}_\pm$ to the boundary components $S_\pm^1$, such that $\partial {\cal D}_\pm$ is the circle $S_\pm^1$ with the opposite orientation; namely,
\EQ{
\Sigma'= \Sigma \cup {\cal D}_+\cup {\cal D}_- \,,\qquad \partial\Sigma'=0\,.
}
We shall consider boundary conditions corresponding to the configurations of minimal energy discussed in section~\ref{SSSG} so that, on $\partial\Sigma$, the field $\gamma$ takes values in conjugacy classes of $H=H^\zeta \times H_\text{ss}$ which are of the form~\eqref{coexp}.
Namely,
\EQ{
\gamma(t,\pm L)= e^{2\pi  \lambda_\pm^\zeta /\kappa}\, g (t,\pm L)\, e^{2\pi i \Blambda_\pm\cdot\Bh /\kappa}\, g^{-1} (t,\pm L)\,,
\label{bc2}
}
where $\lambda^\zeta_\pm$ are constant elements of $\mh^\zeta$, $\Blambda_\pm\cdot\Bh$ are constant elements of a Cartan subalgebra of $\mh_\text{ss}$, and $g(t,\pm L)\in H_\text{ss}$.
Since the conjugacy classes are simply connected, each field $\gamma$ with these boundary conditions can be extended to $\gamma':\Sigma'\to G$ in such a way that $\gamma'({\cal D}_\pm)\in H^\zeta  \times {\cal C}_{\Blambda_\pm}(H_\text{ss})$.
The crucial observation is that the restriction of $\omega^\text{WZ}(\gamma)$ to a coadjoint orbit  is a total derivative~\footnote{In more precise terms, $\omega^\text{WZ}(\gamma)$ is the pullback of the canonical 3-form on $G$ by the field $\gamma$ which, restricted to a coadjoint orbit, becomes exact:
\EQ{
\omega^\text{WZ}(\gamma)\,d^3x\equiv \omega^\text{WZ}(\gamma) \,,\qquad
\omega^\text{WZ}(ge^{2\pi i\lambda/k} g^{-1})= d\alpha^\lambda(g)\,,\qquad
\alpha^\lambda(g)\equiv \epsilon^{ab}\alpha_{ab}^\lambda(g)\, d^2x \,,\qquad
\,.
} 
}
\EQ{
&\omega^\text{WZ}(ge^{2\pi i \Blambda\cdot\Bh /k}g^{-1})= \epsilon^{abc}\,\partial_c  \alpha_{ab}^\Blambda(g) \,,\\[5pt]
&
\alpha_{ab}^\Blambda(g)= 3\Tr\left(e^{-2\pi i \Blambda\cdot\Bh /k}\, g^{-1}\partial_a g \,e^{2\pi i \Blambda\cdot\Bh /k}\, g^{-1}\partial_b g\right)\,.
\label{aaa}
}
Then, following~\cite{Alekseev:1998mc,Gawedzki:1999bq,Gawedzki:2001ye}, the Wess-Zumino-Witten action of the field $\gamma$ with boundary conditions~\eqref{bc2} is defined by
\EQ{
S_\Sigma[\gamma]=\frac{\kappa}{8\pi} \left[\int_\Sigma d^2x\, L^\sigma(\gamma) +\frac{2}{3}\left( \int_B d^3 y\, \omega^\text{WZ}(\tilde{\gamma}') -\sum_{n=\pm}\int_{{\cal D}_n} d^2z\,\epsilon^{ab}\,\alpha^{\Blambda_n}_{ab}(g_n) \right)\right]\,,
\label{WZWb}
}
where $B$ is a three-manifold bounded by $\Sigma'$, $\tilde{\gamma}'$ is an extension of $\gamma'$ from $\Sigma'$ to $B$, and $g_\pm\in H_\text{ss}$ are the corresponding extensions of $g(t,\pm L)$ from $\partial\Sigma$ to ${\cal D}_\pm$ so that
\EQ{
\gamma\big|_{H_\text{ss}}= g_\pm  \, e^{2\pi i \Blambda_\pm\cdot\Bh /\kappa}\, g_\pm^{-1} \quad\text{on}\quad {\cal D}_\pm\,.
\label{bc3}
}
Compared to~\eqref{WZW}, the role of the additional term, which is non-trivial only if $H_\text{ss}\not=\emptyset$, is just to compensate the variation of $\omega^\text{WZ}$ on $\Sigma'-\Sigma={\cal D}_+\cup {\cal D}_-$ so that the equations of motion do not change.

The action $S_\Sigma$ is ambiguously defined because it depends on the choice of the two extensions $\gamma'$  and $\tilde \gamma'$.
This ambiguity does not affect the classical equations of motion, but at the quantum level the path integral measure $\exp(iS_\text{gWZW})$ has to be independent of the choice of $\gamma'$  and $\tilde \gamma'$. This requires that the improved WZ term be uniquely defined modulo $2\pi{\mathbb Z}$ for each $\gamma$ and imposes a quantization condition on $\Blambda_\pm$, in addition to the well known quantization of the coupling constant $\kappa$ summarized by~\eqref{Level}. 
As explained in~\cite{Alekseev:1998mc,Gawedzki:1999bq}, the WZW action~\eqref{WZWb} is well defined modulo $2\pi$ if $\kappa$ is an integer and $\Blambda_\pm$ are dominant (integral) weights $\Blambda_\pm\in P^+$ of level $\kappa$, which are those that satisfy the conditions $\Blambda\cdot \Balpha_i\in {\mathbb Z}\geq0$ for all the simple roots $\Balpha_i$ of $H_\text{ss}$, and $\Blambda\cdot\Btheta\leq \kappa$ for the highest root $\Btheta$.
Compared to~\eqref{clms}, this results in the following {\em quantum moduli space}
\EQ{
{\cal M}_{\text{q}}= \{\Blambda\in P^+\mid 0\leq \Blambda\cdot\Btheta \leq \kappa\}
\label{qms}
}
which labels the conjugacy classes where the boundary conditions of 
$\gamma$ are allowed to take values.

\subsection{The boundary terms}

Using~\eqref{WZWb} for $S_\Sigma$ in~\eqref{gWZW} provides an action that is explicitly invariant under the gauge transformations~\eqref{gauge} generated by $h\in H_\text{ss}$ supplemented by
\EQ{
g_\pm \to h g_\pm\,,
\label{gaugeU}
}
which follows by consistency with~\eqref{bc3}.
In contrast, it is not invariant under the gauge transformations generated by the elements in the abelian subgroup $H^\zeta$. Namely, since the additional term in~\eqref{WZWb} is blind to the gauge transformations generated by $h=e^u\in H^\zeta $ and $\partial B\not=\Sigma$, the gauged WZW action~\eqref{gWZW} transforms as
\EQ{
S_\text{gWZW}[\gamma,A_\mu]\to S_\text{gWZW}[\gamma,A_\mu] - \frac{\kappa}{2\pi}\int_{\Sigma} d^2x\, \epsilon^{\mu\nu}\partial_\mu  \Tr\big(\phi\, \partial_\nu u\big)
\label{noinv}
}
where we have defined $\gamma=e^\phi$. 
We will fix this by adding a boundary term
\EQ{
\widetilde{S}_{\text{gWZW}}[\gamma,A_\mu] = S_{\text{gWZW}}[\gamma,A_\mu] - \frac{\kappa}{2\pi} \int_{\Sigma} d^2x\, \epsilon^{\mu\nu} \partial_\mu\, \Tr\big(\phi\, A_\nu^\zeta\big)\,,
\label{gWZW2}
}
where $A_\mu^\zeta$ is the component of the gauge field $A_\mu$ in $\mh^\zeta$. On general grounds, terms of this form have been considered in~\cite{FigueroaO'Farrill:2005uz}.
The change of the action~\eqref{ala} defined with $\widetilde{S}_\text{gWZW}$ under infinitesimal variations of the fields reads
\EQ{
&\delta S[\gamma,A_\mu]=-\frac{\kappa}{\pi}\int_\Sigma d^2x\, \Big[\Tr\Big(\delta A_+\, D_-\gamma\gamma^{-1} - \delta A_-\, \gamma^{-1}D_+\gamma\big)\\[5pt]
&\quad
+\gamma^{-1}\delta\gamma\, [\partial_++\gamma^{-1}\partial_+\gamma+\gamma^{-1}A_+\gamma-z\mu \Lambda,\partial_-+A_- -z^{-1}\mu\gamma^{-1}\Lambda\gamma] \Big)\\[5pt]
&\quad
+\Tr\Big(\partial_+\left(\delta gg^{-1}\, D_-\gamma\gamma^{-1}\right)-\partial_-\left(\delta gg^{-1}\, \gamma^{-1}D_+\gamma\right)\Big)\bigg]\\[5pt]
&\quad
-\frac{\kappa}{2\pi} \int_\Sigma d^2x \, \epsilon^{\mu\nu}\partial_\mu\Tr\big(\phi\,  \delta A_\nu^\zeta\big)
-\frac{\kappa}{4\pi} \int_\Sigma d^2x \, \partial_\mu\Tr\big(\delta\phi^\zeta\,  \partial^\mu \phi\big)\,.
\label{gwzweom2}
}

Notice that the anomalous term in~\eqref{noinv} and the boundary term in~\eqref{gWZW2} would vanish if we enforce the boundary conditions satisfied by the components of $\gamma$ and $A_\mu$ in $H^\zeta$ and $\mh^\zeta$, respectively,
\EQ{
\gamma^\zeta\big|_{\partial\Sigma}= f^\zeta\,,\qquad
A_\mu^\zeta\big|_{\partial\Sigma}=-\partial_\mu v\,,
\label{bccentre}
}
where $f^\zeta$ is constant. At this point it is important to stress that the definition of the WZ term and, thus, of $\widetilde{S}_{\text{gWZW}}$ depends explicitly only on the form of the component of $\gamma\big|_{\partial\Sigma}$ in $H_\text{ss}$. In contrast, although we have used that $\gamma\big|_{\partial\Sigma}\in H$, it is completely independent of the form of its component in $H^\zeta$ and of the boundary conditions satisfied by $A_\mu$. 
In order to understand the interplay between the definition of the action and the boundary terms, we have to recall an important point that has to be taken into account when looking at variational principles (for example, see~\cite{Julia:1998ys,Dyer:2008hb}). 
Of course, a necessary condition for the action to be stationary is that
the  fields satisfy the (Euler-Lagrange) equations of motion. However, for the action to be truly stationary, any boundary contributions arising
from the variation must vanish, and no conditions other than those provided by the boundary conditions and the equations of motion themselves may be used in checking whether those boundary
contributions vanish. 
In our case, the first two boundary contributions in~\eqref{gwzweom2} vanish  making use of either the constraints~\eqref{gco3} or the boundary conditions, since they correspond to configurations of minimal energy given by covariantly constant group elements of $H$ and, hence, $D_\mu\gamma\big|_{\partial\Sigma}=0$. 
The third and fourth boundary contributions also vanish taking into account~\eqref{bccentre}.

The boundary conditions satisfied by the gauge field $A_\mu$ motivate the introduction of an extra boundary term that becomes crucial to make connection with the semiclassical quantization of the soliton spectrum worked out in~\cite{Hollowood:2010dt,Hollowood:2011fm}.
Recall that the construction of~\eqref{WZWb} is based on the fact that the restriction of the Wess-Zumino term to a conjugacy class is a total derivative. By means of~\eqref{aaa}, this provides $\alpha_{ab}^\Blambda(g)$ which is uniquely defined only up to a total derivative
\EQ{
\alpha_{ab}^\lambda\to \alpha_{ab}^\lambda+\partial_a \psi_b\,.
}
On the other hand, it is not difficult to check that the restriction of the Wess-Zumino term to configurations of the form $\gamma=g\, e^{2\pi i \tilde{\Blambda}\cdot\Bh/\kappa}\, g^{-1}$ with $\tilde{\Blambda}$ not being constant is also a total derivative
\EQ{
&\omega^\text{WZ}(g e^{2\pi i \tilde{\Blambda}\cdot\Bh/\kappa} g^{-1})= \epsilon^{abc}\partial_c  \widetilde\alpha_{ab}^{\tilde{\Blambda}}(g) \,,\\[5pt] 
&
\tilde\alpha_{ab}^{\tilde{\Blambda}}(g)=\alpha_{ab}^{\tilde{\Blambda}}(g)+\frac{12\pi i}{\kappa}\, \Tr\left(\tilde{\Blambda}\cdot\Bh\,  g^{-1}\partial_a g \, g^{-1}\partial_b g\right)\,.
\label{bbb}
}
For $\tilde{\Blambda}=\Blambda$ constant, this motivates the following non-minimal choice of $\alpha_{ab}^\Blambda$
\EQ{
\tilde\alpha_{ab}^{\Blambda}(g)=\alpha_{ab}^{\Blambda}(g)-\frac{12\pi i}{\kappa}\,\partial_a \Tr\left(\Blambda\cdot\Bh\,  g^{-1}D_b g\right)\,.
\label{aaaB}
}
Compared to~\eqref{bbb}, we have changed $g^{-1}\partial_b g$ into the gauge invariant combination $g^{-1}D_b g$ that involves the covariant derivative $D_\mu g=(\partial_\mu +A_\mu^\text{ss})g$, where $A_\mu^\text{ss}$ is the component of the gauge field $A_\mu$ in $\mh_\text{ss}$.
Using $\gamma=e^\phi$ and the form of the component of $\gamma\big|_{\partial\Sigma}$ in $H_\text{ss}$ given by~\eqref{bc2},
\EQ{
\gamma\big|_{H_\text{ss}}= ge^{2\pi i \Blambda_\pm\cdot\Bh/\kappa}g^{-1}\;\Rightarrow\;
\frac{2\pi i}{\kappa}\, \Blambda_\pm\cdot\Bh
=  g^{-1}\,\phi g\big|_{\mh_\text{ss}} \quad \text{at}\quad x=\pm L\,,
}
which leads to our final proposal for the (bosonic part of the) Lagrangian action of the generalized SG theories subject to the boundary conditions~\eqref{bc1}
\EQ{
&S[\gamma,A_\mu]=
-\frac\kappa{2\pi}\int_\Sigma d^2x\,\Tr\,\Big[
\gamma^{-1}\partial_+\gamma\,\gamma^{-1}\partial_-\gamma + 2\mu^2 \Lambda \gamma^{-1} \Lambda \gamma\\[5pt]
&
\quad+2\big(A_+\partial_-\gamma\gamma^{-1}
-A_-\gamma^{-1}\partial_+\gamma-\gamma^{-1}A_+\gamma A_-+A_+A_-\big) 
+ \epsilon^{\mu\nu} \partial_\mu\, \big(\phi\, A_\nu^\zeta\big)\Big]
\\[5pt]
&
\quad
+\frac{\kappa}{12\pi}\left[\int_B d^3 y\, \omega^\text{WZ}(\tilde{\gamma}') -\sum_{n=\pm}\int_{{\cal D}_n} d^2z\,\epsilon^{ab}\,\Big(\alpha^{\lambda_n}_{ab}(g_n)-6\partial_a \Tr(\phi\, D_b g_ng_n^{-1})\Big)\right]\,.
}
In order to validate it, we have to check that the boundary contributions generated by the extra term actually vanish using the boundary conditions. Compared to~\eqref{gwzweom2}, they are of the form
\EQ{
\delta S[\gamma,A_\mu]
\to
\delta S[\gamma,A_\mu]
-\frac{\kappa}{2\pi} \sum_{n=\pm}\int_{{\cal D}_n} d^2z\,\epsilon^{ab}\,\partial_a \Tr\Big(\phi\, \delta\big(D_b g_ng_n^{-1}\big)
+[\phi, D_bg_ng_n^{-1}]\, \delta g_ng_n^{-1}\Big)\,.
\label{gwzweom4}
}
Since $\partial{\cal D}_+\cup \partial{\cal D}_-=-\partial\Sigma$ 
and $g_\pm$ are the extensions of $g(t,\pm L)$ from $\partial\Sigma$ to ${\cal D}_\pm$, the two new boundary contributions vanish if we impose
\EQ{
D_\mu gg^{-1}\big|_{\partial\Sigma}=0
}
which is equivalent to $A_\mu^\text{ss}\big|_{\partial\Sigma}=-\partial_\mu gg^{-1}$. Namely, if the gauge field is flat on the boundary, which is the boundary condition~\eqref{bc1} satisfied by $A_\mu^\text{ss}$.

Since $D_bg_ng_n^{-1}=\partial_bg_ng_n^{-1} + A_b^\text{ss}$ and ${\cal D}_+\cup{\cal D}_+=- \partial\Sigma$, we can write the action as
\EQ{
&S[\gamma,A_\mu]=
-\frac\kappa{2\pi}\int_\Sigma d^2x\,\Tr\,\Big[
\gamma^{-1}\partial_+\gamma\,\gamma^{-1}\partial_-\gamma + 2\mu^2 \Lambda \gamma^{-1} \Lambda \gamma\\[5pt]
&
\quad+2\big(A_+\partial_-\gamma\gamma^{-1}
-A_-\gamma^{-1}\partial_+\gamma-\gamma^{-1}A_+\gamma A_-+A_+A_-\big) 
+ \epsilon^{\mu\nu} \partial_\mu\, \big(\phi\, A_\nu\big)\Big]
\\[5pt]
&
\quad
+\frac{\kappa}{12\pi}\left[\int_B d^3 y\, \omega^\text{WZ}(\tilde{\gamma}') -\sum_{n=\pm}\int_{{\cal D}_n} d^2z\,\epsilon^{ab}\,\Big(\alpha^{\lambda_n}_{ab}(g_n)-6\partial_a \Tr(\phi\, \partial_b g_ng_n^{-1})\Big)\right]
\label{ala2}
}
which clarifies the origin of the $A_\mu$-dependent boundary term introduced in~\cite{Hollowood:2010dt,Hollowood:2011fm,Hollowood:2011fq}.
It is worth noticing that, taking into account the explicit dependence of $\alpha_{ab}^\Blambda$ on the coupling constant $\kappa$, the two last terms in~\eqref{ala2} provide contributions of higher order in perturbation theory at order $1/\kappa$. It would be interesting to investigate the role of these terms in perturbation theory particularly in the light of the 
puzzles that appear in such calculations \cite{Hoare:2010fb} (see also \cite{Hoare:2013ysa}).

\section{The Conserved Noether Charges}
\label{noether}

The main purpose of this section is to provide a Lagrangian interpretation for the spin-zero charge $q_0$ as a Noether charge and to identify the underlying global symmetry transformations.
The gauge invariant definition of conserved charges in non-abelian gauge theories has been extensively discussed in the literature. For instance, it is interesting to look at~\cite{Abbott:1982jh}, where the gauge invariant charges of magnetic monopoles and dyons are deduced as particular examples. A more thorough discussion, with many references, can be found in~\cite{Julia:1998ys,Silva:1998ii} which we briefly summarize in the next subsection for the sake of completeness.

\subsection{Some generalities}

The starting point is the extension of Noether's theorem to local symmetries, which was discussed by Noether herself and by Hilbert. It has two important consequences. The first one is that the naive Noether current is locally exact modulo the equations of motion. The second,  is that the local symmetry actually gives rise to an infinite number of conserved currents which are also locally exact. However, only a subset of them, singled out by the boundary conditions, gives rise to conserved quantities.

Consider a Lagrangian action $S=\int L(\varphi,\partial\varphi)$ that is invariant under a continuous global field transformation $\delta\varphi=\epsilon\,\Delta(\varphi)$
\EQ{
\delta S=0 \Leftrightarrow \delta L  = \left[\frac{\partial L}{\partial \varphi} -\partial_\mu\left(\frac{\partial L}{\partial(\partial_\mu\varphi)}\right)\right] \delta\varphi +\partial_\mu \left[\frac{\partial L}{\partial(\partial_\mu\varphi)}\ \delta\varphi\right]= \epsilon\,\partial_\mu R^\mu\,.
\label{inv}
}
This implies the existence of a conserved current $J^\mu$
\EQ{
J^\mu = \frac{\partial L}{\partial(\partial_\mu\varphi)}\ \Delta(\varphi) - R^\mu,\qquad
\partial_\mu J^\mu =-\frac{\delta L}{\delta \varphi}\, \Delta(\varphi)\approx 0\,,
\label{NC}
}
where 
\EQ{
\frac{\delta L}{\delta \varphi} \equiv \frac{\partial L}{\partial \varphi} -\partial_\mu\left(\frac{\partial L}{\partial(\partial_\mu\varphi)}\right)\approx 0
}
are the equations of motion, and the notation $\approx$ indicates equality on-shell. 
This is the well known Noether's theorem. It is useful to recall that the Noether current can be constructed by looking at the local version of the global symmetry transformation with $\epsilon=\epsilon(x)$ which, in general, is not a symmetry of the Lagrangian action. Then
\EQ{
\delta L&= \epsilon\left[\frac{\delta L}{\delta \varphi}\, \Delta(\varphi) + \partial_\mu \left(\frac{\partial L}{\partial(\partial_\mu\varphi)}\ \Delta(\varphi)\right)\right] + \partial_\mu\epsilon\,\left(\frac{\partial L}{\partial(\partial_\mu\varphi)}\ \Delta(\varphi)\right)\\[5pt]
&=\partial_\mu \epsilon \, J^\mu + \partial_\mu\left(\epsilon R^\mu\right)\,
}
and the Noether current is provided (off-shell) by the coefficient of $\partial_\mu \epsilon$.

If the symmetry transformation is local, $\delta\varphi$ can be written in terms of a local parameter ${u}^a(x)$ and its derivatives. In the usual cases, we have
\EQ{
\delta \varphi = {u}^a \Delta_a(\varphi) + \partial_\nu{u}^a \Delta_a^\nu(\varphi)\,.
}
$R^\mu$ can be expanded in a similar way but, for simplicity, we will assume that it vanishes. Then, expanding~\eqref{inv} in terms of ${u}^a$ and its derivatives and using their arbitrariness, it can be easily shown that
\EQ{
&J_a^\mu = \frac{\partial L}{\partial(\partial_\mu\varphi)} \Delta_a(\varphi) =\partial_\nu U_a^{\mu\nu}- \frac{\delta L}{\delta\varphi}\, \Delta^\nu_a(\varphi) \approx \partial_\nu U_a^{\mu\nu}, \\[5pt]
&U_a^{\mu\nu} =-U_a^{\nu\mu} = \frac{\partial L}{\partial(\partial_\mu\varphi)} \,\Delta_a^\nu(\varphi)\,.
\label{Exact}
}
In other words, the Noether current $J_a^\mu$ corresponding the global transformation associated to ${u}^a$ is locally exact (topological) modulo the equations of motion. It is worth noticing that $U_a^{\mu\nu}$ is defined off-shell.

The second consequence of the theorems of Noether and Hilbert follows by considering local transformations along a fixed direction; namely,
\EQ{
{u}^a(x) = \epsilon (x)\> \xi_0^a(x),
}
where $\epsilon(x)$ is the local parameter for the abelian subgroup of transformations generated by $\xi^a_0(x)$. 
In this case,
\EQ{
\delta\varphi= \epsilon\big( \Delta_a(\varphi)\xi_0^a +\Delta_a^\nu(\varphi) \partial_\nu\xi_0^a\big) +\partial_\nu\epsilon\, \big(\Delta_a^\nu(\varphi)\,\xi_0^a \big)\,,
}
and it can be shown that the corresponding Noether current is also exact
\EQ{
J^\mu_{\xi_0} \approx \partial_\nu U_{\xi_0}^{\mu\nu},\qquad 
U_{\xi_0}^{\mu\nu}= U_a^{\mu\nu}\xi^a_0=-U_{\xi_0}^{\nu\mu}\,.
\label{exact}
}
Therefore, considering all the possible choices of $\xi^a_0(x)$ we conclude that the local symmetry actually gives rise to an infinite number of conserved currents, and that all of them are topological modulo the equations of motion.
Most of these currents do not lead to conserved charges. As explained in~\cite{Abbott:1982jh} and~\cite{Julia:1998ys,Silva:1998ii}, conserved charges are associated to the generators of gauge transformations that leave the boundary conditions invariant (boundary Killing vectors).

\subsection{Noether charges}

The infinitesimal form of the gauge transformations~\eqref{gauge} reads
\EQ{
\delta \gamma = [u,\gamma]\,,\qquad \delta A_\mu= [u,A_\mu]-\partial_\mu u\,,\qquad u\in\mh
}
which, if we expand $u$ in terms of a basis of $\mh$ as $u=u^a T_a$, corresponds to
\EQ{
\Delta_a(\gamma)= [T_a,\gamma]\,,\qquad
\Delta_a(A_\mu)= [T_a,A_\mu]\,,\qquad
\Delta_a^\nu(A_\mu)=- \delta_\mu^\nu\, T_a\,.
}
Let $\xi_0=\xi_0(x)$ be a fixed function taking values in $\mh$ and consider the global transformation corresponding to $u(x)=\epsilon\, \xi_0(x)$.
Using~\eqref{exact}, the action~\eqref{ala2} gives rise to the Noether current
\EQ{
J_{\xi_0}^\mu \approx \partial_\nu U^{\mu\nu}_{\xi_0}\,,\qquad 
U^{\mu\nu}_{\xi_0}= \Tr\left[\frac{\partial S}{\partial(\partial_\mu A_\rho)}
\,\Delta_a^\nu(A_\rho)\,\xi_0^a\right]= \frac{\kappa}{2\pi}\, \epsilon^{\mu\nu} \Tr\big(\phi\, \xi_0\big)
\label{NCgen}
}
which is topological as expected.

Conserved charges are obtained with specific choices of $\xi_0$ singled out by the boundary conditions. Take
\EQ{
Q_{\xi_0}=\int_{-\infty}^{+\infty} dx\, J_{\xi_0}^0 \approx -\frac{\kappa}{2\pi}\, \Tr\big(\phi\, \xi_0\big)\Big|_{x=-\infty}^{x=+\infty}
\label{Nch}
}
and consider the boundary conditions~\eqref{bc1}. Then, $Q_{\xi_0}$ is conserved provided that 
\EQ{
\partial_0\big(U^{-1}\xi_0 U\big)\big|_B= 0\,,\qquad
A_\mu\big|_B=-\partial_\mu UU^{-1}\big|_B
\label{cond}
}
which means that $\xi_0$ is the generator of an infinitesimal gauge transformation that leaves $A_0$ invariant on the boundary. In other words, $\xi_0$ satisfies the boundary Killing equation $D_0\xi_0\big|_B=0$~\cite{Julia:1998ys,Silva:1998ii,Abbott:1982jh}.

The simplest solutions of~\eqref{cond} are provided by $\xi_0\in \mh^\zeta$ constant, which shows that the global gauge transformations generated by the elements of $H^\zeta$ give rise to conserved quantities. Moreover, they are gauge invariant and defined off-shell.

Solutions of~\eqref{cond} taking values in $\mh_\text{ss}$ can be constructed on-shell so that the Noether charge $Q_{\xi_0}$ is related to the spin-zero charge given by~\eqref{q0}. 
As explained in sections~\ref{SSSG} and~\ref{lagrangian}, in~\eqref{bcgood} the component of $f_\pm$ in $H_\text{ss}$ takes values in conjugacy classes of $H$ of the form 
\EQ{
f_\pm = g_\pm e^{2\pi i \Blambda_\pm\cdot \Bh/\kappa} g_\pm^{-1} \;\Rightarrow \; 
\phi \big|_{x=\pm \infty}= O(1/\kappa)\,.
}
Then, eq.~\eqref{q0} can be linearized and simplifies to
\EQ{
e^{q_0} -1\simeq  -U^{-1}(\infty)\phi(\infty) U(\infty)+ U^{-1}(-\infty)\phi(-\infty) U(-\infty) + O(1/\kappa^2)\,.
}
Here $\phi(\pm\infty)\equiv \phi(t,\pm\infty)$, and it is worth recalling that $U(\pm\infty)$ is a simplified notation for $U((t,\pm\infty);x_0)$, where $x_0$ is the arbitrary reference point introduced in~\eqref{Udef}. Then, for
\EQ{
\xi_0(x)= U(x;x_0) \, v\, U^{-1}(x;x_0)
\label{choice}
}
with $v\in\mh$ constant, the charge~\eqref{Nch} reads
\EQ{
Q_{\xi_0} \equiv Q[v]\simeq \frac{\kappa}{2\pi} \Tr\big(v\, q_0\big)+ O(1/\kappa)
\label{NC2}
}
which provides the interpretation of $q_0$ as a Noether charge in the semiclassical, $\kappa\to\infty$, limit.
Notice that $\xi_0(x) \equiv \xi_0(x;x_0;v)$
whose transformation under the gauge symmetry~\eqref{gauge} is
\EQ{
\xi_0(x;x_0;v) \to h(x)\, \xi_0(x;x_0; h^{-1}(x_0) v h(x_0))\, h^{-1}(x)\,,
\label{choice2}
}
which is consistent with the transformation properties of $q_0$ given by~\eqref{qtransf}. Therefore, both $Q[v]$ and $q_0$ are invariant under $H^\zeta$ gauge transformations, and under $H_\text{ss}$-gauge transformations that satisfy $h(x_0)=1$.

Eq.~\eqref{choice} only makes sense on-shell because it involves the Wilson line defined in~\eqref{Udef} using that $A_\mu$ is flat. However, the general discussion of the previous subsection suggests that it should be possible to define the exact conserved current~\eqref{NCgen} off-shell. 
We can do it by imposing the off-shell gauge fixing condition $A_0=0$, which is consistent with the boundary conditions. The corresponding residual gauge transformations 
are of the form~\eqref{gauge} with $\partial_0 h=0$ which, restricted to our (time-like) boundary, look like global gauge transformations. Thus, the remaining component of the gauge field can be written off-shell as
\EQ{
A_1= -\partial_x \widehat{U}\, \widehat{U}^{-1}\,,\qquad
\widehat{U}(t,x)=\text{P}\,\exp\Big[-\int_{s_0}^{x} ds\, A_1(t,s)\Big]\equiv \widehat{U}(x;s_0)
\label{Uhatdef}
}
where $s_0$ is a reference value for $x$. Then, the off-shell quantity
\EQ{
\widehat{\xi}_0= \widehat{U}(x;s_0) \,v\, \widehat{U}^{-1}(x;s_0)
} 
with $v\in\mh$ constant provides a conserved Noether charge $Q_{\widehat{\xi}_0}\equiv \widehat{Q}[v]$.
Since it is constructed using a particular gauge fixing prescription, this off-shell charge is obviously not gauge invariant. However, we can check that, on-shell, its value coincides with $Q[v]$. First of all, recall that the on-shell charge $Q[v]$ is invariant under $H_\text{ss}$-gauge transformations that satisfy $h(x_0)=1$. If $x_0=(t_0,s_0)$, the group valued function $\widetilde{h}$ defined by
\EQ{
\widetilde{h}^{-1}(t,x)=\text{P}\,\exp\Big[-\int_{t_0}^{t} d\tau\, A_0(\tau,x)\Big] 
}
generates a gauge transformation that takes $A_0\to 0$. Correspondingly, the on-shell Wilson line transforms as $U(t,x;x_0)\to \widehat{U}(x;s_0)$ and, since $\widetilde{h}(t_0,s_0)=1$, we conclude that
\EQ{
\widehat{Q}[v]\approx Q[v]\,.
}
As we have already alluded to, we expect that the non-local nature of these charges in the classical theory will lead to them satisfying a quantum group deformation of the Lie algebra in the quantum theory~\cite{Bernard:1990ys}.

\subsection{Global gauge transformations and conserved charges}

We will deduce the interpretation of $q_0$ as a conserved Noether charge in a slightly different way to show that the underlying global symmetry transformation is a specific combination of global and local gauge transformations.
Let us
consider the following composition of local (l) and global (g) transformations
\EQ{
&
\gamma \buildrel (l)\over{\hbox to 17pt{\rightarrowfill}} r\gamma r^{-1}\buildrel (g)\over{\hbox to 17pt{\rightarrowfill}} hr\gamma r^{-1}h^{-1}\buildrel (l)\over{\hbox to 17pt{\rightarrowfill}} r^{-1}\big(hr\gamma r^{-1}h^{-1}\big) r\,,\\[5pt]
&
A_\mu \buildrel (l)\over{\hbox to 17pt{\rightarrowfill}} r(A_\mu+\partial_\mu)r^{-1}
\buildrel (g)\over{\hbox to 17pt{\rightarrowfill}} hr(A_\mu+\partial_\mu)r^{-1}h^{-1}
\buildrel (l)\over{\hbox to 17pt{\rightarrowfill}} r^{-1} \big(hr(A_\mu+\partial_\mu)r^{-1}h^{-1}+\partial_\mu\big) r
}
where $r=r(t,x)\in H$ and $h$ is constant. This is just the gauge transformation generated by $r^{-1}hr$, which is an obvious symmetry of the action.
The corresponding Noether current can be found following standard means by considering the local version of the transformation with $h=h(t,x)$ so that
\EQ{
&\gamma\to \widehat \gamma=(r^{-1}hr)\,\gamma\, (r^{-1}hr)^{-1}\,,\\[5pt]
&A_\mu\to \widehat{A}_\mu =(r^{-1}hr)\big(A_\mu + r^{-1} h^{-1}\partial_\mu h r+\partial_\mu\big)  (r^{-1}hr)^{-1}\,.
}
Gauge invariance is the statement that 
\SP{
S[\widehat \gamma,\widehat{A}_\mu] =S[\gamma,A_\mu +r^{-1}h^{-1}\partial_\mu h\, r ]\,.
\label{NotGlobal}
}
Finally, for an infinitesimal transformation $h\simeq 1+u$, and to linear order in $u$,
\SP{
\delta_uS&=S[\widehat \gamma,\widehat{A}_\mu] -S[\gamma,A_\mu]\simeq-\frac{\kappa}{\pi} \int_\Sigma d^2x \Big[\Tr\big(\partial_+ u\, r(D_-\gamma\gamma^{-1})r^{-1} -\partial_-u\, r(\gamma^{-1}D_+\gamma)r^{-1} \big)\\[5pt]
&
\hspace{6.5cm}+\frac{1}{2}\, \epsilon^{\mu\nu}\partial_\mu \Tr\big(r^{-1}\phi r\,\partial_\nu u\big)\Big]\,,
\label{InfGaugeb}
}
where the last term comes from the $A_\mu$-dependent boundary term in~\eqref{ala2}.
Using the equations of motion~\eqref{gco3}, this leads to the on-shell expression for the Noether current
\EQ{
{\cal J}^\mu \approx   \epsilon^{\mu\nu} \partial_\nu (r^{-1}\phi r)
\label{currb}
}
which is related to~\eqref{NCgen} as follows
\EQ{
J_{\xi_0}^\mu =\frac{\kappa}{2\pi}\, \Tr\big( v {\cal J}^\mu\big)\,,\qquad \xi_0=rvr^{-1}
}
where $v\in\mh$ is constant. In the previous section we showed that this current gives rise to a conserved quantity for $v\in \mh^\zeta$, which means that the current is independent of $r$ and corresponds to the global gauge transformations generated by $H^\zeta$.
For generic choices of $v\in \mh_\text{ss}$, this current gives rise to a conserved charge provided that
\EQ{
\partial_0(r^{-1} U)\big|_B=0\,,\qquad A_\mu\big|_B=-\partial_\mu UU^{-1}\big|_B\,.
}
Then, the relevant symmetry is a global gauge transformation acting on the gauge slice singled out by $A_0\big|_B=0$, which matches the off-shell definition of the charges proposed in the previous subsection.

In~\cite{Hollowood:2010dt}, using the on-shell gauge fixing condition $A_\mu=0$, it was shown that global gauge transformations act on the internal moduli space of soliton solutions. Consider a soliton solution $\gamma^s=\gamma^s(t,x)$ with $A^s_\mu=0$. Eq.~\eqref{NotGlobal} allows one to calculate the change of the action under the transformation $\gamma_s\to h(t) \gamma^s h^{-1}(t)$, which provides the effective action used it that reference to perform the semiclassical quantization of the soliton spectrum. 
First of all, since $A^s_\mu=-\partial_\mu UU^{-1}=0$, then $U\big|_{x=\pm \infty}\equiv U_\pm$ are constant elements of $H$ and, therefore, $\phi\big|_{x=\pm \infty}$ are constant too. Then, using~\eqref{NotGlobal},
\SP{
S[h(t) \gamma^s h^{-1}(t),0]=S[\gamma^s,0]+\frac{\kappa}{2\pi} \int dt\, \Tr\Big( h^{-1} \frac{dh}{dt}\, \sigma\Big) +\cdots
\label{effact}
}
with
\EQ{
\sigma= -\phi\big|_{x=+ \infty} +\phi\big|_{x=- \infty}\,,
}
which reproduces the eq.~(6.5) of~\cite{Hollowood:2010dt}. 

\subsection{Global axial transformations generated by $H^\zeta$}
\label{axial}

On a world-sheet without boundary, the action~\eqref{gWZW} is also invariant under the global (axial) transformations
\EQ{
\gamma\to h\gamma h\,,\qquad A_\mu\to A_\mu\,,\qquad h=e^u\in H^\zeta \,,
\label{axialtrans}
}
which correspond to $\phi^\zeta\to \phi^\zeta + 2u$. However, the boundary terms induce the following non-trivial change of the action
\EQ{
\delta S = -\frac{2\kappa}{\pi} \int_\Sigma d^2x\, \Tr\big( u\, F_{+-}\big) = -\frac{\kappa}{\pi} \int_{\partial\Sigma} dx^\mu \, \Tr\big( u\, A_\mu^\zeta\big)\,.
\label{anomaly}
}
This agrees with the familiar statement that if we gauge a vector $\U(1)$, then the axial $\U(1)$ is anomalous~\cite{Maldacena:2001ky}. In other words, the axial $H^\zeta $ symmetry is broken, and only the discrete subgroup singled out by the condition $e^{i\delta S}=1$ is non anomalous and provides a good symmetry of the theory.

In our case, the world-sheet is given by~\eqref{ws} and $\partial\Sigma$ consists of two timeline circles $S_T^1$ located at $x=\pm L$ with $L\to\infty$. Then, the anomalous contribution~\eqref{anomaly} can be written as
\EQ{
\delta S = -\frac{\kappa}{\pi} \int_{S_T^1} dt \, \Tr\big( u\, A_0^\zeta\big)\Big|_{x=-L}^{x=+L}\;.
\label{anomaly2}
}
With our boundary conditions~\eqref{bc1}, the gauge field on $\partial\Sigma$ is exact
\EQ{
A_0\big|_{x=\pm L}= -\partial_0 U\, U^ {-1}\big|_{x=\pm L}\,.
\label{period1}
}
Moreover, both $\gamma$ and $A_\mu$ satisfy periodic boundary conditions in~$t$
\EQ{
\gamma(t+T,x)=\gamma(t,x)\,,\qquad
A_\mu(t+T,x)=A_\mu(t,x)
}
which are preserved by the gauge transformations~\eqref{gauge} provided that
they satisfy
\EQ{
h(t+T,x)h^{-1}(t,x) \in H\cap \text{Cent}(G)\,.
\label{centG}
}
In other words, we have to consider topologically non-trivial gauge transformations whose generators are periodic up to the elements of the centre of $G$. For flat gauge fields, like those in the boundary, we can write $A_\mu =-\partial_\mu UU^{-1}$ and normalize $U(x_0)=1$ like in section~\ref{cc}. Then, the behaviour of $U$ under gauge transformations is
\EQ{
U(t,x)\to U^h(t,x)= h(t,x) U(t,x) h^{-1}(x_0)
}
so that
\EQ{
U^h(t+T,x)\, U^{h\, -1}(t,x) = h(t+T,x) U(t+T,x) U^{-1}(t,x) h(t,x)\,.
}
Notice that the periodicity of $A_\mu$ implies that $U(t+T,x) U^{-1}(t,x)$ is independent of $t$ and $x$ and, taking~\eqref{centG} into account, we conclude that
\EQ{
U(t+T,x)\, U^{-1}(t,x) \in \text{Cent}(G)\cap H\,.
}
Therefore, the component of $A_0$ in $H^\zeta$ on the boundary satisfies
\EQ{
A_0^\zeta\big|_{x=\pm L}= -\partial_0\, \omega_\pm\,,\qquad \omega_\pm (t+T)-\omega_\pm (t) = 2\pi \widetilde{Y}_\pm
}
where $\widetilde{Y}_\pm$ are generators of $\mh^\zeta$ such that $e^{2\pi \widetilde{Y}_\pm}\in \text{Cent}(G)$.
Then,
\EQ{
\delta S = -2\kappa \, \Tr\big( u\, (\widetilde{Y}_+-\widetilde{Y}_-)\big)\,,
\label{anomaly3}
}
and the condition $e^{i\delta S}=1$ is equivalent to
$\kappa\Tr\big( u\, \widetilde{Y}_\pm\big)/\pi\in {\mathbb Z}$.
Since \mbox{$\phi^\zeta\to \phi^\zeta + 2u$}, this motivates the following quantization condition for $\phi^\zeta\big|_{\partial\Sigma}$
\EQ{
\frac{\kappa}{2\pi}\, \Tr\big(\phi^\zeta\ \widetilde{Y}\big)\big|_{\partial\Sigma}\in {\mathbb Z}
\,.
\label{qaxial}
}
for any $\widetilde{Y}\in\mh^\zeta$ such that $e^{2\pi \widetilde{Y}}\in \text{Cent}(G)$

\section{Generalized SG Soliton Kinks}
\label{solitons}

In this section we will apply the quantization conditions of the boundary conditions summarized by~\eqref{qms} and~\eqref{qaxial} to the soliton solutions constructed in~\cite{Hollowood:2010dt} using the gauge fixing condition $A_\mu=0$. To be specific, most of the discussion will be restricted to the solitons of the theories associated to the complex projective spaces $\CP^{n+1}=\SU(n+2)/\U(n+1)$. The generalization to other cases is straightforward  although technically involved.

The construction of~\cite{Hollowood:2010dt} makes use of the eigenvalues and eigenvectors of $\Lambda$, which can be written as
\EQ{
\Lambda = \sum_{a=1}^p im_a \Bv_a  \Bv_a^\dagger\,,\qquad \Lambda \Bv_a=im_a \Bv_a\,,
\label{lg}
}
where the maximal number of linearly independent non-vanishing eigenvalues equals the rank of the symmetric space. 
Then, the solitons are associated to two eigenvectors $\Bv_a,\Bv_b$ corresponding to different eigenvalues $m_a\not=m_b$ and are labelled by a real number $0<q<\pi$. If $m_a<m_b$, the asymptotic values of the function $\chi=\chi(z)$ introduced in~\eqref{LPsol1} (with $g(z)=1$) are
\SP{
&\log \chi(z)\big|_{x=+\infty}=\sum_{\sigma\in\II}
\log\left[\frac{z-\sigma_i(\xi)^*}{z-\sigma_i(\xi)}\right]\sigma\left(\Bv_{a}\Bv_{a}^\dagger\right)
\,,\\[5pt]
&\log\chi(z)\big|_{x=-\infty}=\sum_{\sigma\in\II}
\log\left[\frac{z-\sigma_i(\xi)^*}{z-\sigma(\xi)}\right]\sigma\left(\Bv_{b}\Bv_{b}^\dagger\right)
\,,
\label{asval}
}
which determine the subtracted monodromy and the value of the conserved charges. In~\cite{Hollowood:2010dt}, the group $F$ is thought of as a subgroup of $\SU(n_F)$ where $n_F$ is the dimension of the defining representation of $F$. It is
picked out as the invariant subgroup of an involution $\sigma_+$ so that $\II=\{\sigma_-,\sigma_+\}$, where $\sigma_-$ is the involution whose stabilizer is $G$. 
It is worth noticing that this construction actually realizes $F$ as a subgroup of $\U(n_F)$ and, consequently, it provides the field $\gamma$ only up to a compensating factor needed to ensure that $\text{det\,}\gamma=1$. The form of this factor follows by looking at the 
associated linear problem~\eqref{alp} satisfied by $\Upsilon=\chi\Upsilon_0$ (with $A_\mu=0$)
\AL{
\partial_+\chi(z) \chi^{-1}(z) + z\chi(z) \Lambda\chi^{-1}(z)& = -
\gamma^{-1}\partial_+\gamma+z\Lambda\ ,
\label{Eq1}\\[5pt]
\partial_-\chi(z) \chi^{-1}(z) + z^{-1}\chi(z) \Lambda\chi^{-1}(z)& = z^{-1}\gamma^{-1}\Lambda\gamma\ .
\label{Eq2}
}
Identifying the residues of the both sides of~\eqref{Eq1} and ~\eqref{Eq2} at $z=\infty$ and $z=0$, respectively, we find
\EQ{
\chi(\infty)\Lambda\chi^{-1}(\infty) =\Lambda\,,\qquad
\chi(0) \Lambda \chi^{-1}(0)=\gamma^{-1}\Lambda\gamma\, .
\label{NormalDF}
}
The first condition is solved by $\chi(\infty)=1$, while the second gives
\EQ{
\gamma= \psi\chi^{-1}(0)\,,\qquad \psi^{-1}\Lambda \psi=\Lambda
\label{reconstruction}
}
where $\psi\in \U(n_F)$ is constant and satisfies $\sigma_-(\psi)=\sigma_+(\psi)=\psi$.
Consequently, the boundary conditions of the solitons are given by
\SP{
&\phi_+\equiv \phi\big|_{x=+\infty}=-2q\,\sum_{\sigma\in\II}
\sigma \left(i\Bv_{a}\Bv_{a}^\dagger- ...
\right)
\,,\\[5pt]
&\phi_-\equiv\phi\big|_{x=-\infty}=-2q\,\sum_{\sigma\in\II}
\sigma \left(i\Bv_{b}\Bv_{b}^\dagger- ...
\right)
\label{asympold}
}
where the ellipsis represents a common term that commutes with $\Lambda$ needed to ensure that $\Tr\phi_\pm=0$. Their mass is
\EQ{
M=\frac{4\kappa}{\pi}\big|(m_b-m_a)\sin q\big|
}
and, since $[\phi_+,\phi_-]=0$, they carry the spin-zero charge
\SP{
q_0=-\phi_++\phi_- =2q\,\sum_{\sigma\in\II}
\sigma \left(i\Bv_{a}\Bv_{a}^\dagger- i\Bv_{b}\Bv_{b}^\dagger
\right)
}
which is independent of the choice of $\psi$ in~\eqref{reconstruction}.

For $F/G=\SU(n+2)/\U(n+1)$, we will use the fundamental representation of $\SU(n+2)$ which consists of matrices acting on the vector space spanned by the orthonormal vectors $\Be_a$, with $a=1,\ldots,n+2$. The elements of $G= \U(n+1)$ are of the form
\EQ{
\left(\begin{array}{c|c}e^{i\varphi/(n+1)}\, {\mathbb I}_{n+1} & 0 \\\hline 0 & e^{-i\varphi}\end{array}\right)\cdot \left(\begin{array}{c|c}M & 0 \\\hline 0 & 1\end{array}\right)\,,\qquad \varphi\in{\mathbb R}\,,\qquad M\in \SU(n+1)
}
which is invariant under $\varphi\to\varphi - 2\pi p$ and $M\to e^{2\pi i p/(n+1)}M$ with $p\in{\mathbb Z}$. This exhibits that \mbox{$G=(\U(1)\times \SU(n+1))/{\mathbb Z}_{n+1}$}. The constant element $\Lambda$ is given by
\EQ{
\Lambda=m\left(E_{n+2,n+1}-E_{n+1,n+2}\right)
\label{lambda}
}
which has three eigenvalues: $\{+im,-im,0\}$. The non-null ones are non-degenerate and their eigenvectors come in pairs
\EQ{
\Lambda\Bv_\pm=\pm im \Bv_\pm\ ,\qquad\Bv_\pm=\frac1{\sqrt{2}}\big(\Be_{n+1}\mp i\Be_{n+2}\big)\,.
\label{eige}
}
In contrast, the null eigenvalue is $n$-times degenerate. A basis of eigenvectors is provided by $\Be_a$ for $a=1,\ldots,n$ so that a generic null eigenvector is a linear combination of the form $\BOmega=\sum_{a=1}^{n} c_a \Be_a$ with complex coefficients. In the following $\BOmega$ will always denote a generic null eigenvector normalized such that $\BOmega^\dagger\BOmega=1$.
The elements of $H=\U(n)$ are of the form
\EQ{
\left(\begin{array}{c|c}e^{2i\varphi/n}\, {\mathbb I}_{n} & 0 \\\hline 0 & e^{-i\varphi}\,{\mathbb I}_{2}\end{array}\right)\cdot 
\left(\begin{array}{c|c}C & 0 \\\hline 0 & {\mathbb I}_{2}\end{array}\right)\,,\qquad \varphi\in{\mathbb R}\,,\qquad C\in \SU(n)
\label{hform}
}
whch is invariant under $\varphi\to\varphi - 2\pi p$ and $C\to e^{4\pi i p/n}C$ and exhibits that $H=(\U(1)\times \SU(n))/{\mathbb Z}_{n}$.

The elementary $\CP^{n+1}$ solitons, those that cannot be split into more elementary ones, are associated to $m_a=0$ and $m_b=m$ (and to $m_a=-m$ and $m_b=0$). 
In this case, in~\eqref{reconstruction}, we can choose 
\EQ{
\psi=\left(\begin{array}{c|c}e^{4iq/n}\,{\mathbb I}_{n} & 0 \\\hline 0 & {\mathbb I}_2\end{array}\right)
\label{gchoice}
}
which commutes with $\Lambda$,
so that their boundary conditions are given by
\SP{
&\phi_+=-2iq\, \left(2\BOmega\BOmega^\dagger- \frac{2}{n}\; {\mathbb I}_{n}
\right)=-2iq\,U\left(2\Be_1\Be_1^\dagger- \frac{2}{n}\; {\mathbb I}_{n}
\right)U^\dagger
\,,\\[5pt]
&\phi_-=-2iq\,\left(\Bv_{+}\Bv_{+}^\dagger+ \Bv_{-}\Bv_{-}^\dagger-\frac{2}{n}\; {\mathbb I}_{n}
\right)= -2iq\,\left(E_{n+1,n+1} + E_{n+2,n+2}-\frac{2}{n}\; {\mathbb I}_{n}
\right)\,,
\label{ab}
} 
where we have used that $\BOmega$ can be written as
$\BOmega = U \Be_1$
with $U\in H=\U(n)$. However, the solitons are not sensitive to the overall phase of $\BOmega$ and, hence, we can restrict $U\in \SU(n)$.
Therefore, the boundary value $\phi_-$ takes values in $\mh^\zeta=\muu(1)$ whose infinitesimal generator will be normalized as
\EQ{
Y=i\Big(E_{n+1,n+1}+E_{n+2,n+2} - \frac{2}n\,{\mathbb I}_{n}\Big)\,.
}
In contrast, $\phi_+$ takes values in $\mh_\text{ss}=\msu(n)$.
The positive step operators of $\msu(n)$ are $\Be_i \Be_j^\dagger$ with $1\leq i<j\leq n$ so that
\EQ{
[\Be_1\Be_1^\dagger- \frac{1}{n}\, {\mathbb I}_{n}, \Be_i \Be_j^\dagger]= \delta_{i,1} \Be_i \Be_j^\dagger
}
and, therefore,
\EQ{
\Be_1\Be_1^\dagger- \frac{1}{n}\, {\mathbb I}_{n}= \Bomega_1\cdot\Bh
}
where $\Bomega_1$ is the first fundamental weight. Then, we can write the boundary values~\eqref{ab} as
\AL{
&\phi_+= U\left(-4q i\, \Bomega_1\cdot\Bh \right) U^\dagger=
\widetilde U\left(+4q i\, \Bomega_{n-1}\cdot\Bh \right)\widetilde U^\dagger\in \mh_\text{ss}=\msu(n)\,,
\label{fiplusa}\\[5pt]
&\phi_-=-2q\, Y\in \mh^\zeta =\muu(1)\,,
\label{fiplusb}}
where we have used that the weights $+\Bomega_1$ and $-\Bomega_{n-1}$ are related by means of a Weyl transformation.
It is worth noticing that the spin-zero charge
\EQ{
q_0 =-\phi_+ +\phi_- =U\left[ 2iq \left(2\Be_1\Be_1^\dagger - E_{n+1,n+1}- E_{n+2,n+2}\right)\right] U^\dagger\,,
}
as well as the whole subtracted monodromy,
is actually ambiguous up to shifts $q\to q + \pi$ since it only appears via $e^{q_0}$, which confirms that the inequivalent solitons correspond to $0<q<\pi$. However, this is not true for the boundary conditions satisfied by $\gamma$
\EQ{
q\to q+\pi \;\Rightarrow\; &e^{\phi_\pm}\to  
\left(\begin{array}{c|c}e^{4\pi i/n}{\mathbb I}_n & 0 \\\hline 0 & {\mathbb I}_2\end{array}\right)\, e^{\phi_\pm}
=e^{-4\pi i\, \Bomega_1\cdot\Bh}\, e^{\phi_\pm}=e^{-2\pi \, Y}\, e^{\phi_\pm}\,.
}
This is in agreement with the fact that 
$H=(\U(1)\times \SU(n))/{\mathbb Z}_{n}$ with ${\mathbb Z}_{n}$ diagonally embedded in $H^\zeta=\U(1)$ and $H_\text{ss}=\SU(n)$. In other words, although the elements of $\mh$ can be uniquely decomposed in terms of their components in $\mh_\text{ss}$ and $\mh^\zeta$, the decomposition of the elements of $H$ as a product of a component in $H_\text{ss}$ and a component in $H^\zeta$ is unique only up to multiplication by an element of ${\mathbb Z}_n= H_\text{ss}\cap H^\zeta$, which is the centre of $H_\text{ss}$.
With this caveat, eq.~\eqref{fiplusb} shows that $\gamma(-\infty)=e^{\phi_-}$ takes values in $H^\zeta=U(1)$, and eq.~\eqref{fiplusa} shows that 
$\gamma(+\infty)=e^{\phi_+}$ takes values in a conjugacy class of $H_\text{ss}=\SU(n)$ associated to $-q\Bomega_1$ or, equivalently, to $+q\Bomega_{n-1}$.

According to~\eqref{Level}, for $\CP^{n+1}$ the coupling constant $\kappa=k$ is a positive integer. Then, if we write $e^{\phi_+}= Ue^{2\pi i\Blambda_+\cdot\Bh/k}U^\dagger$, the different conjugacy classes are labelled by $\Blambda_+$ in the classical moduli space ${\cal M}_\text{cl}$ given by~\eqref{clms}. 
It is
convenient to label the solitons using
\EQ{
\overline{q}= \begin{cases}
q     &,\;\; 0<q\leq \frac\pi2, \\[5pt]
q-\pi &,\;\; \frac\pi2\leq q<\pi
\end{cases}
}
instead of $q$, so that $-\frac\pi2\leq \overline{q}\leq \frac\pi2$. Then,
\EQ{
e^{\phi_+}= \begin{cases}
\widetilde{U}\,  e^{4\overline{q}i\Bomega_{n-1}\cdot \Bh}\, \widetilde{U}^\dagger
\;\Rightarrow\; \Blambda_+=\frac{2k\overline{q}}{\pi}\, \Bomega_{n-1}\in {\cal M}_\text{cl}   &,\;\; 0<\overline{q}\leq \frac\pi2, \\[5pt]
e^{-2\pi Y}\, U\,  e^{4|\overline{q}|i\Bomega_1\cdot \Bh}\, U^\dagger
\;\Rightarrow\; \Blambda_+=\frac{2k|\overline{q}|}{\pi}\, \Bomega_1\in {\cal M}_\text{cl}  &,\;\; - \frac\pi2\leq \overline{q}<0\,,
\end{cases}
\label{bcc}
}
and the quantization conditions summarized by $\Blambda_+\in{\cal M}_\text{q}$, with ${\cal M}_\text{q}$ given by~\eqref{qms}, read
\EQ{
\overline{q}=\frac{\pi N}{2k}\,,\qquad N=\pm1,\ldots,\pm k\,.
\label{topoq}
}
These solutions have mass
\EQ{
M=\frac{4k\mu}{\pi}\left|\sin\left(\frac{\pi N}{2k}\right)\right|
}
which agrees with the results of the semiclassical quantization worked out in~\cite{Hollowood:2010dt}.

The quantization of $\phi_+^\text{ss}$ implies the quantization of both $\phi_+^\zeta$ and $\phi_-^\zeta$ and, remarkably, the resulting values satisfy the quantization rule~\eqref{qaxial}.
In this case, $\text{Cent}(G)\cap H$ is the discrete group generated by $e^{2\pi \widetilde{Y}}=e^{-4\pi i/(n+2)}\,{\mathbb I}_{n+2}$ with
\EQ{
\widetilde{Y}= \frac{n}{n+2}\, Y\,.
}
First of all, the components of $\phi_\pm^\zeta$ corresponding to elements in ${\mathbb Z}_n=H_\text{ss}\cap H^\zeta$ satisfy the quantization rule trivially. They are of the form
\EQ{
\phi^\zeta= -p \, 2\pi\, Y \;\Rightarrow\; e^{\phi^\zeta}=e^{-4\pi i p\Bomega_1\cdot\Bh}\,,\qquad p\in{\mathbb Z}\,,
}
so that
\EQ{
\frac{k}{2\pi}\,\Tr(\phi^\zeta \widetilde{Y})= 2kp\in 2k{\mathbb Z}\,.
}
Notice that the components of $\phi_+$ in $\mh^\zeta$ for $-\pi/2\leq \overline{q}<0$ in~\eqref{bcc} are precisely of this form (with $p=1$). The non-trivial check concerns $\phi_-$ which reads
\EQ{
\phi_-=\phi_-^\zeta=-2q\, Y \;\Rightarrow\; \frac{k}{2\pi}\,\Tr(\phi_-^\zeta \widetilde{Y})=\frac{2kq}{\pi}\in {\mathbb Z}
}
as a consequence of~\eqref{topoq}. 
According to the discussion in section~\ref{axial}, this reflects the breakdown of the anomalous axial symmetry generated by $H^\zeta=\U(1)$ to a discrete one associated to the subgroup ${\mathbb Z}_{2nk}$.

Looking at $\phi_\pm^\text{ss}$ (for $n\geq2$), the resulting picture is that the solitons of~\cite{Hollowood:2010dt} are kinks that interpolate between a discrete number of conjugacy classes of $H_\text{ss}=\SU(n)$ described in terms of the dominant weights of  level $\leq k$. Namely, for $m_a=0$ and $m_b=m$
\EQ{
0<q\leq\frac\pi2\quad \longrightarrow\quad  (\phi_-^\text{ss},\phi_+^\text{ss})&= (0, N\Bomega_{n-1})\equiv K_{0, N\Bomega_{n-1}}\,,\\[5pt]
\frac\pi2\leq q< \pi \quad \longrightarrow\quad  (\phi_-^\text{ss},\phi_+^\text{ss})&= (0, N\Bomega_1)\equiv K_{0, N\Bomega_1}\,,}
while for $m_a=-m$ and $m_b=0$
\EQ{
0<q\leq\frac\pi2\quad \longrightarrow\quad  (\phi_-^\text{ss},\phi_+^\text{ss})&= (N\Bomega_{n-1},0)\equiv K_{N\Bomega_{n-1},0}\,,\\[5pt]
\frac\pi2\leq q< \pi \quad \longrightarrow\quad (\phi_-^\text{ss},\phi_+^\text{ss})&= (N\Bomega_1,0)\equiv K_{N\Bomega_1,0}\,,
}
with $N=1,\ldots,k$,
and it is natural to think of the latter as the anti-solitons of the former.

All this  fits nicely the kink picture used in~\cite{Hollowood:2010rv} to construct the S-matrix. However, it is important to notice that this construction only provides solitons with only either $\phi_+$ or $\phi_-$ having components in $\mh_\text{ss}=\msu(n)$. 
In contrast, both the quantization conditions summarized by~\eqref{qms} and the kink picture used in~\cite{Hollowood:2010rv}
suggest that there should exists a more general class of soliton solutions where, in particular, both $\phi_+$ and~$\phi_-$ have components in $\mh_\text{ss}=\msu(n)$.
It is not difficult to generalize the construction of~\cite{Hollowood:2010dt} to produce a larger class of soliton solutions with boundary conditions of this type. The key observation is that the equations of motion with $A_\mu=0$, which read
\EQ{
\partial_-\big(\gamma^{-1}\partial_+\gamma\big)=\mu^2\big[\Lambda\,,\, \gamma^{-1}\Lambda\gamma\big]\,,\qquad
\big(\gamma^{-1}\partial_+\gamma\big)^\perp=\big(\partial_-\gamma\gamma^{-1}\big)^\perp=0\,,
}
are invariant under the global transformation
\EQ{
\gamma(t,x)\to h_L \gamma(t,x)\,,\qquad h_L\in H\,,
\label{GT}
}
in addition to the (residual) global gauge transformations.
Even though this is not a symmetry of the action, it can be applied to the soliton solutions of~\cite{Hollowood:2010dt} to construct new ones with different asymptotic values
\EQ{
e^{\phi_\pm} \to e^{\phi'_\pm}= h_Le^{\phi_\pm}\,.
\label{soltrans}
}
Notice that the subtracted monodromy~\eqref{mono} does not change under this transformation, which means that all the conserved charges, and in particular $q_0$, remain invariant.
In fact, this amounts to changing the compensating factor $\psi$ in~\eqref{reconstruction} as $\psi\to h_L \psi$.
This can be used to fix the form of the boundary conditions at, say, $x=+\infty$ as follows
\EQ{
\gamma(t,x)\to \widehat\gamma(t,x)= \big(g_+ e^{2\pi i\lambda_+/k}g_+^{-1}\big)\, e^{-\phi_+}\, \gamma(t,x)
\label{newsol}
}
where $\lambda_+=\lambda_+^\zeta+ \Blambda_+\cdot\Bh$, with $\lambda_+^\zeta$ a constant element of $\mh^\zeta$, $\Blambda_+\cdot\Bh$ a constant element of the Cartan subalgebra of $\mh_\text{ss}$, and $g_+\in H_\text{ss}$ in agreement with~\eqref{bc2}. Then, taking into account the form of the boundary conditions at $x=-\infty$,
\EQ{
\widehat\gamma(t,-\infty)= \big(g_+ e^{2\pi i\lambda_+/k}g_+^{-1}\big)\, e^{-\phi_+}\,e^{+\phi_-}=\big(g_- e^{2\pi i\lambda_-/k}g_-^{-1}\big)\,.
\label{proA}
}
The resulting relation between the spin-zero charge $q_0$ and the boundary conditions is
\EQ{
e^{q_0} =\big(g_+ e^{-2\pi i\lambda_+/k}g_+^{-1}\big)\,\big(g_- e^{2\pi i\lambda_-/k}g_-^{-1}\big)\,.
\label{proB}
} 
This means that $e^{q_0}$, which lives in a conjugacy class itself, takes values in the product of the conjugacy classes corresponding to the boundary conditions at $x=\pm\infty$.
               
The new solitons are kinks that interpolate between conjugacy classes of $H$ labelled by $\lambda_+$ and $\lambda_-$. However, for generic conjugacy classes, the topological quantization of the boundary conditions does not imply the quantization of the component of $q_0$ in $\mh_\text{ss}$. This can be explicitly checked for $H=\SU(2)$ using the results of~\cite{Jeffrey:1992,Jeffrey:2005}. 
Let us denote
by $C(\lambda)$ the conjugacy class of the matrix $\text{diag}(e^{i\pi \lambda/k},e^{-i\pi \lambda/k})$ with $0\leq\lambda\leq k$. Then, one can solve for $g\in \SU(2)$ in 
\EQ{
g\left(\begin{array}{cc}e^{i\pi \lambda_3/k} & 0 \\0 & e^{-i\pi\lambda_3/k}\end{array}\right) g^{-1} \in C(\lambda_1)C(\lambda_2)
}
provided that $\lambda_3$ is any real number $0\leq \lambda_3\leq k$ such that
\EQ{
|\lambda_1-\lambda_2|\leq \lambda_3 \leq \text{min}\big\{\lambda_1+\lambda_2, 2k-(\lambda_1+\lambda_2)\big\}\,.
\label{pretensor}
}

The quantization of the component of $q_0$ in $\mh_\text{ss}$ can be established following the approach of~\cite{Hollowood:2010dt} to quantize the internal degrees-of-freedom of the solitons in the semiclassical approximation. The soliton solutions with asymptotic values~\eqref{asval} carry an internal collective coordinate corresponding to the vector $\Bvarpi=\Bv_a+\Bv_b$. It gives rise to an internal classical moduli space that can be identified with the orbit
\EQ{
\Bvarpi\to U\Bvarpi\,,\qquad U\in H_\text{ss}\,,
\label{intmod}
}
which is equivalent to the global gauge transformation $\gamma\to U\gamma U^{-1}$. Then, the semi-classical quantization of the soliton makes use of an effective finite dimensional Lagrangian constructed by allowing the collective coordinates to become time-dependent taking $U\to U(t)$ in~\eqref{intmod} or, equivalently, by substituting $\gamma\to U(t)\gamma U^{-1}(t)$ into the action of the theory.
By construction, the new soliton solutions~\eqref{newsol} also carry the internal collective coordinate $\Bvarpi=\Bv_a+\Bv_b$. However, in this case the transformation~\eqref{intmod} corresponds to a global gauge transformation supplemented by $g_+\to U^{-1} g_+$. The latter is just a change of the particular point in the conjugacy class where $\widehat{\gamma}$ takes values at $x=+\infty$ which does not change the value of the conserved charges. Therefore, we can still follow the conventional route to the semi-classical quantizaton of the solution by substituting 
$\gamma\to U(t)\gamma U^{-1}(t)$ into the action to find the effective quantum mechanical system of the collective coordinate $U(t)$. It is given by eq.~\eqref{effact}
\SP{
S[U(t)\, \widehat{\gamma}\, U^{-1}(t),0]=S[\widehat{\gamma},0]+\frac{k}{2\pi} \int dt\, \Tr\Big( U^{-1} \frac{dU}{dt}\, \sigma\Big) +\cdots
}
where, using~\eqref{proA},
\EQ{
\sigma= -\widehat{\phi}_+ +\widehat{\phi}_-= q_0
}
and $q_0$ actually means its component in $\mh_\text{ss}$. For $\CP^{n+1}$, $q_0$ is given by~\eqref{fiplusa} and~\eqref{fiplusb}; namely,
\EQ{
q_0 =U\Big(4i\,|\overline{q}|\, \Bomega_1\cdot\Bh\Big) U^\dagger
\quad\text{or}\quad
\widetilde U\Big(4i\,|\overline{q}|\, \Bomega_{n-1}\cdot\Bh\Big) \widetilde U^\dagger
}
with $0< |\overline{q}|\leq \frac\pi2$, and the results of~\cite{Hollowood:2010dt} imply that  $\overline{q}$ is quantized precisely as in~\eqref{topoq}.
Once $q_0$ is quantized, as well as $\lambda_\pm$, notice that~\eqref{pretensor} applied to~\eqref{proB} (for $n=2$) is remarkably reminiscent of the truncated tensor product recently considered in~\cite{Hoare:2013ysa} in the context of the $q$-deformed $\text{AdS}_5\times S^5$ string S-matrix for $q$~a root of unity. 

\section{Conclusions}
\label{end}

The main purpose of this paper has been to show that the kink picture used in~\cite{Hollowood:2010rv} and~\cite{Hoare:2013ysa} to construct the S-matrix of the generalized SG theories arises in a natural way from their Lagrangian formulation. 
We have performed a detailed construction of the (bosonic part of the) Lagrangian action $S[\gamma,A_\mu]$ of these theories, which is provided by the gauged Wess-Zumino-Witten action for a coset $G/H$ deformed by a specific potential term. Since it includes a Wess-Zumino topological term, its form depends on the boundary conditions satisfied by the field $\gamma$, and its consistency imposes quantization conditions on them, in addition to the well known quantization of the coupling constant. 
More precisely, since $H$ is the product of a semi simple Lie group $H_\text{ss}$ and an abelian group $H^\zeta=U(1)^{\times p}$, the consistency of the WZ term imposes quantization conditions on the boundary conditions satisfied by the components of $\gamma$ in $H_\text{ss}$. We have also argued that the quantization of the boundary conditions satisfied by the components of $\gamma$ in $H^\zeta$ is a consequence of the breakdown of a global (axial) symmetry generated by $H^\zeta$ which becomes a discrete symmetry.
Applied to soliton solutions, the resulting picture is that they are kinks that interpolate between a discrete set of vacua represented by conjugacy classes of the symmetry group $H$, which fits nicely the kink picture used in those articles.

Even though this will be discussed in detail in the follow up paper~\cite{Nextpap}, we would like to finish by pointing out that the correspondence between the Lagrangian formulation and the $S$-matrix kink picture goes beyond the quantization of the soliton boundary conditions. 
In the semiclassical limit, $\kappa\to\infty$, the vacuum configurations represented by the conjugacy classes can be related to the quasi-classical states (coherent states) introduced in~\cite{Hollowood:2010rv}. Then, the relationship between the spin-zero (Noether) charge $q_0$ and the boundary conditions provided by~\eqref{proB} can be understood as a semi-classical realization of a $q$-deformed Clebsch-Gordan decomposition.
This is in agreement with the expectation that, as a consequence of their non-local nature, the conserved charges of these theories satisfy a quantum group deformation of the Lie algebra of the symmetry group rather than the conventional Lie algebra.

\section*{Acknowledgements}

\noindent
TJH is supported by the STFC grant ST/G000506/1.

\noindent
JLM is supported in part by MINECO (FPA2011-22594), the Spanish Consolider-Ingenio 2010 Programme CPAN (CSD2007-00042), and FEDER. 

\noindent
DMS is supported by FAPESP. He also thanks Xunta de Galicia for support while this work was in progress and IGFAE for hospitality.

\end{document}